\documentclass[prd,twocolumn,tightenlines,preprintnumbers,showpacs,superscriptaddress,notitlepage,nofootinbib,eqsecnum,floatfix,longbibliography,aps,10pt]{revtex4-1}
\pdfoutput=1
\usepackage[utf8]{inputenc}
\usepackage[T1]{fontenc}
\usepackage{mathrsfs}
\usepackage{bbold}
\usepackage{amsmath}
\usepackage{amssymb}
\usepackage{mathtools}
\usepackage{kbordermatrix}
\usepackage{amsfonts,dsfont}
\usepackage{array}
\usepackage{bm,bbm}
\usepackage{graphicx}
\usepackage{xcolor}
\usepackage{enumitem}
\usepackage{soul}
\usepackage{stmaryrd}
\usepackage{hyperref}
\usepackage{cancel}
\usepackage{tikzsymbols}
\usepackage{diagbox}
\usepackage[normalem]{ulem}
\hypersetup{
    colorlinks=true,     % false: boxed links; true: colored links
    linkcolor=blue,      % color of internal links
    citecolor=blue,      % color of links to bibliography
    filecolor=blue,      % color of file links
    urlcolor=blue        % color of external links
}
\DeclareMathOperator{\sign}{sign}

\newcommand{\G}{\mathscr{G}}

\newcommand{\id}{\mathbb{1}}

\newcommand{\sx}{\sigma^x}
\newcommand{\sy}{\sigma^y}
\newcommand{\sz}{\sigma^z}
\newcommand{\eqnref}[1]{Eq.~(\ref{#1})}
\newcommand{\figref}[1]{Fig.~\ref{#1}}
\newcommand{\tabref}[1]{Table~\ref{#1}}
\newcommand*{\ket}[1]{\left|{#1}\right\rangle}

\newcommand\scalemath[2]{\scalebox{#1}{\mbox{\ensuremath{\displaystyle #2}}}}

\makeatletter
\def\l@subsubsection#1#2{}
\makeatother

\begin{document}
\title{Improving Schr\"odinger Equation Implementations with Gray Code for Adiabatic Quantum Computers}

\author{Chia~Cheng~Chang}
\email{chiachang@berkeley.edu}
\affiliation{RIKEN iTHEMS, Wako, Saitama 351-0198, Japan}
\affiliation{Department of Physics, University of California, Berkeley, California 94720, USA}
\affiliation{Nuclear Science Division, Lawrence Berkeley National Laboratory, Berkeley, California 94720, USA}
\affiliation{LinkedIn Corporation, Sunnyvale, California 94085, USA}
\author{Kenneth S. McElvain}
\affiliation{Department of Physics, University of California, Berkeley, California 94720, USA}
\affiliation{Nuclear Science Division, Lawrence Berkeley National Laboratory, Berkeley, California 94720, USA}
\author{Ermal Rrapaj}
\affiliation{Department of Physics, University of California, Berkeley, California 94720, USA}
\author{Yantao Wu}
\affiliation{RIKEN iTHEMS, Wako, Saitama 351-0198, Japan}
\affiliation{Department of Physics, University of California, Berkeley, California 94720, USA}
\affiliation{Department of Physics, Princeton University, Princeton, New Jersey 08544, USA}

\newcommand{\alert}[1]{\textbf{\color{red}{#1}}}
\renewcommand{\vec}[1]{\boldsymbol{#1}}

\begin{abstract}
  We reformulate the continuous space Schr\"odinger equation in terms of spin Hamiltonians. For the kinetic energy operator, the critical concept facilitating the reduction in model complexity is the idea of position encoding. A binary encoding of position produces a spin-$1/2$ Heisenberg-like model and yields exponential improvement in space complexity when compared to classical computing. Encoding with a binary reflected Gray code (BRGC), and a Hamming distance 2 Gray code (H2GC) reduces the model complexity down to the XZ and transverse Ising model respectively.
For $A$ qubits  BRGC  yields $2^A$ positions and is reduced to its 2-local form with $\mathrm{O}(A)$ ancillary qubits. H2GC yields $2^{A/2 + 1}$ positions with $\mathrm{O}(A^2)$ 3-local penalty terms.
We also identify the bijective mapping between diagonal unitaries and the Walsh series, producing the mapping of any real potential to a series of $k$-local Ising models through the fast Walsh transform.
Finally, in a finite volume, we provide some numerical evidence to support the claim that the total time needed for adiabatic evolution is protected by the infrared cutoff of the system. 
As a result, initial state preparation from a free-field wavefunction to an interacting system is expected to exhibit polynomial time complexity with volume and constant scaling with respect to lattice discretization for all encodings. For H2GC, if the evolution starts with the transverse Hamiltonian due to hardware restrictions, then penalties are dynamically introduced such that the low lying spectrum reproduces the energy levels of the Laplacian. The adiabatic evolution of the penalty Hamiltonian is therefore sensitive to the ultraviolet scale. It is expected to exhibit polynomial time complexity with lattice discretization, or exponential time complexity with respect to the number of qubits given a fixed volume. 
\end{abstract}

\preprint{RIKEN-iTHEMS-Report-22, N3AS-21-003}

\maketitle
%%%%%%%%%%%%%%%%%%%%%%%%%%%%%%%%%
%%%%%%%%%%%%%%%%%%%%%%%%%%%%%%%%%
%%%%%%%%%%%%%%%%%%%%%%%%%%%%%%%%%
%\tableofcontents

%========================================================================================
\section{INTRODUCTION}
\label{sec:introduction}
%========================================================================================
\newcommand{\norm}[1]{\left\lVert#1\right\rVert}

The understanding of many physical problems requires obtaining Schr\"odinger equation solutions for the system under study.
In this work, we develop techniques to solve it with adiabatic quantum computing.
A typical classical computing choice for numerically solving the Schr\"odinger equation is to pick a discrete basis in which to express the Hamiltonian. Then, one diagonalizes the resulting matrix, either completely, or for very large problems, uses techniques such as the Lanczos algorithm to find low-lying eigenstates and eigenvalues.
The discrete basis can, for example, be comprised of the states of a harmonic oscillator, or some other exactly solvable Hamiltonian.  Other useful basis choices are a discrete position, or momentum basis.
As a first step, we focus on a simple version of the problem: a one-body system with a local potential in a $D$-dimension periodic position basis.
\begin{equation}
	-\frac{\hbar^2}{2 M} \nabla^2 \psi(x)+ V(x)\psi(x) = E \psi(x).
\end{equation}
For simplicity of notation, throughout this paper, we work in natural units, $\hbar=c=1$.
We discretize the equation on a lattice with spacing $a$ and $N$ positions in each of the $D$ directions.
Then, up to a discretization error proportional to $a$, the Laplacian becomes an $N \times N$ matrix which acts on the discretized wave function $\psi(a\mathbf{m})$:
\begin{equation}
  \begin{split}
  \nabla^2 \psi(x) &\approx  \frac{1}{a^2}(L\psi)(a\mathbf{n}) \\
  &= \frac{1}{a^2}\left[ \left( \sum\limits_{\mathbf{m} \in \mathcal{N}(n)} \psi(a\mathbf{m})\right) - 2D \psi(a\mathbf{n})\right] ,
\end{split}
\end{equation}
with $\mathcal{N}(\mathbf{n})$ indicating the set of immediate neighbors of the discrete point $\mathbf{n}$.
Here, we use $L$ to denote the dimensionless part of the Laplacian.
The discrete Schr\"odinger equation then follows as
\begin{equation}
	(H\psi) (a\mathbf{n}) = -\frac{1}{2 M a^2}  (L\psi)(a\mathbf{n}) + V(a\mathbf{n})\psi(a\mathbf{n}) = E \psi(a\mathbf{n}).
\end{equation}

The next step is to encode the positions in states of qubits (spins).
One choice of encoding in use is to associate position $i$ with a set of qubits~\cite{Abel2021,Pilon2021}.  In each position, the value of the function is given by a fixed-point representation of the qubits. The advantage of such an encoding is that the solution is diagonal in the computational basis, and is implementable with quantum annealers available today. However, the number of qubits is comparable to the number of classical bits required to solve the same problem.

Alternatively, one can associate positions with $A$-body qubit states in the computational basis and identify
each basis state's amplitude with the wave function at the corresponding point.
Such an association produces a lattice with $2^A$ sites, yielding an exponential improvement in space complexity.
This approach appears in circuit-based quantum algorithms~\cite{Mocz2021},
specifically associating the state of qubit $i$ with the value of bit $i$ of the position index.
Bit $i$ of a number is the coefficient of $2^i$ in the base-2 representation of the number.

In this work, we explore the advantages of other encoding possibilities which yield simpler spin Hamiltonians.
A first encoding choice uses the binary reflected Gray code (BRGC) to represent the sequence of positions, with the bits of the code having the same connection to the qubit states as before.
This option requires only the  $\sigma_i^x\sigma_j^z$ operators, in addition to the $\sigma^x_i$ and $\sigma^z_i\sigma^z_j$ operators in the transverse-field Ising model, and has the key advantage of allowing the Laplacian matrix be reduced to a 2-local form with $\mathrm{O}(A)$ number of auxillary qubits. The BRGC encoding preserves the maximum $2^A$ lattice sites that can be generated from $A$ qubits.
Gray codes are proposed for encoding ladder states in $d$-level systems to simplify raising and lowering operators in gate-based quantum computing \cite{Sawaya2020}.
In \cite{DiMatteo2020} this idea is applied to finding the ground-state energy of a deuteron in a harmonic oscillator basis with a simulated variational quantum eigensolver (VQE). Here, we extend the application of BRGC to map the Schr\"odinger equation, in any dimension, to the $XZ$ model.

A second Gray code, which we call a Hamming-distance-2 Gray code (H2GC), introduces an alternative mapping of the Schr\"odinger equation requiring only the transverse-field Ising model, {\it i.e.} containing only two-body $\sigma^z_i\sigma^z_j$ and one-body $\sigma^x_i$ couplings. The mapping retains an exponential number of valid lattice sites associated with $A$ bit codes in the sequence, while the invalid codes are nulled using an $\mathrm{O}(A^2)$ number of 3-local penalty terms. As a result, the H2GC formulation is polynomially equivalent to BRGC while reducing the complexity of the spin model.

For the examples in this work we consider a single particle moving in $D$ dimensions.
In general, the extension to $m$ particles moving in $d$ spatial dimensions is equivalent to $D = md$.
If each of the $D$ dimensions is discretized into $N$ lattice points,
one needs $\log_2(N^D) = D\log_2(N)$ qubits to represent the entire discretized lattice.

In this work, we treat all particles as being distinguishable, {\it i.e.} particles with Boltzmann statistics.
In quantum chemistry, one often starts from a potential energy surface of atoms and studies the atoms directly without tracking the dynamics of the electrons \cite{Born1927,Car1985}.
In this case, Boltzmann statistics gives an accurate description of molecules when identical particles, {\it e.g.} the hydrogen nuclei in the molecule malonaldehyde $\mathrm{C}_3  \mathrm{H}_4 \mathrm{O}_2$, are not too close in the ground-state wave  function.
For example, the tunneling splitting energy and the quantum momentum distribution can be computed accurately with sampling techniques assuming Boltzmann statistics~\cite{Matyus2016, Wu2020}.
In problems like these, our method offers a direct way to efficiently compute the ground-state wave function of molecules with adiabatic quantum computing.
Particles with bosonic or fermionic statistics are important and natural extensions, but are beyond the scope of this work.

For the convenience of the reader, the notation used throughout this work is defined in Sec.~\ref{subsec:def}.
In Sec.~\ref{sec:laplacian_construction} we provide the mapping of the discretized Laplacian to a $k$-local Hamiltonian,
in binary, BRGC, and H2GC codes in Sec.~\ref{sec:Laplacian:binary}, \ref{sec:Laplacian:Gray}, and \ref{sec:Laplacian:Dist2Gray} respectively. Then, we proceed to describe the mapping of the local potential to any Gray code in Sec.~\ref{sec:potential_decomposition}. Having provided all the necessary steps for encoding the Hamiltonian, in Sec.~\ref{sec:simulations} we provide various simulations of quantum adiabatic computation of the ground state. Specifically, in Sec.~\ref{sec:sim_deuteron} we study a BRGC encoded $S$-wave nucleon potential that reproduces the deuteron binding energy. In Sec.~\ref{sec:sim_quartic} we focus on a two-dimensional quartic and quadratic set of potentials activated in different time intervals to study both initial state preparation and time evolution of the system. We also provide an example of the H2GC code with a harmonic oscillator potential in Sec.~\ref{sec:sim_ho}. We conclude with a summary of our results in Sec.~\ref{sec:summary}.

\subsection{Notation and definitions}
\label{subsec:def}
Before we begin our discussion, the notation used throughout the paper is defined here for clarity.

First, we define a bijection between binary bits and qubit states.
Spin up will be associated with a bit value of 0 or $\ket{0}$, and spin down with a bit value of 1 or $\ket{1}$.
Basis states of an $A$-body qubit system are therefore associated with an $A$-bit binary string with the usual interpretation as an integer in a base-2 representation.
For matrices and vectors over the basis, we order the entries according to the integer value of the corresponding state's bit string.
Let $\id$ be the $2\times2$ identity matrix, and $\sx,\sy,$ and $\sz$ be the Pauli matrices:
\begin{equation}
  \sx = \begin{bmatrix}0&1\\1&0\end{bmatrix}, \,\sy = \begin{bmatrix}0&-i\\i&0\end{bmatrix}, \, \sz = \begin{bmatrix}1 & 0 \\ 0 & -1 \end{bmatrix}
\end{equation}
The set $\{\id, \sx, i\sy, \sz\}$ forms a basis of $2\times2$ real matrices.
Thus, any real matrix of size $2^A\times 2^A$ has a unique tensor product decomposition with these four matrices.

Throughout the paper, for a matrix $M$ of size $2^A\times2^A$, or an array $V$  of size $2^A$, the indices will be denoted by square brackets: $M[k,m]$ is the $(k,m)$ element of $M$ and $V[k]$ is the $k$th element of $V$.
Subscripts of operators denote the qubit index.
When we index qubits, we start from 0 and count from the right.
For example,
\begin{equation}
    \sz_1 = \id \otimes \id \otimes \sz \otimes \id
    \label{eq:sz1}
\end{equation}
means that $\sz$ is acting on qubit 1, while the tensor product of the three     identity operators explicitly states that we are working in a Hilbert-space of four qubits. In the subscript notation, the dimension of the Hilbert space is unspecified, and is explicitly stated if necessary (\textit{e.g.}, when we provide explicit examples).

For multiqubit operators such as the Laplacian, we explicitly list all indices in the subscript. For example, we label a three-qubit Laplacian operator acting on qubits 0, 1, 2 as
\begin{equation}
    L^{(3, \textrm{bin})}_{0\dots 2}.
\end{equation}
Additionally, the superscript ``$\textrm{bin}$'' denotes that the Laplacian is expressed in binary order. In this work, we also derive the Laplacian in ``BRGC'' and ``H2GC'' forms for the binary reflected Gray code and the Hamming-distance-2 Gray code.

For convenience in what follows we define qubit (spin) projection operators
\begin{equation}
  P^0 = \frac{\id + \sz}{2} = \begin{bmatrix}1&0\\0&0\end{bmatrix},\quad  P^1 = \frac{\id - \sz}{2} = \begin{bmatrix}0&0\\0&1\end{bmatrix}
\end{equation}
where $P^0$ projects onto $\ket{0}$ (spin up)  and $P^1$ onto $\ket{1}$ (spin down) for a single qubit.
Raising and lowering operators on a spin are defined as
\begin{align}
  \sigma^+ =& \frac{\sigma^x + i \sigma^y}{2} = \sx P^1 = \begin{bmatrix}0&1\\0&0\end{bmatrix},\nonumber \\
  \sigma^- =& \frac{\sigma^x - i \sigma^y}{2} = \sx P^0 = \begin{bmatrix}0&0\\1&0\end{bmatrix}.
\end{align}

The variable $A$ indicates the number of qubits in the system.

Readers who do not speak binary as a first or second language are highly encouraged to read App.~\ref{sec:math_review}, which summarizes the various binary representations used in this work and their related Walsh functions, which are the foundation of our construction of arbitrary real potentials.
 We make substantial use of Karnaugh maps~\cite{karnaugh1953map} in describing the construction of the H2GC Laplacian.
Karnaugh maps are used in boolean circuit minimization and we include a brief introduction biased towards our application in App.~\ref{sec:KMap}.
Additionally, for readers who would enjoy a more in-depth overview of orthogonal functions and Gray codes, there are many textbooks available in the literature (\textit{e.g.}~\cite{Rao1976}).

\section{CONSTRUCTION OF THE LAPLACIAN}
\label{sec:laplacian_construction}
In this section, we present the mapping of the discrete Laplacian to $k$-local Hamiltonians. The simplest form of the discrete Laplacian is given by the nearest-neighbor finite-difference method,
\begin{equation}
    \frac{\partial^2}{\partial x^2}f(x) = \frac{1}{a^2}\left[f(x+a) + f(x-a) - 2f(x)\right]
\end{equation}
where $a$ is the lattice spacing. In operator form, the dimensionless part of the discrete Laplacian is a tridiagonal matrix with additional nonzero entries in the ends of the antidiagonal due to periodic boundary conditions.
For example, a one-dimensional (1D) lattice with $2^3$ lattice sites has a binary encoded Laplacian operator given by
\begin{equation}
  L^{(3, \textrm{bin})}_{0\dots 2} =
  \begin{bmatrix}
    0& 1& 0& 0& 0& 0& 0& 1 \\
    1& 0& 1& 0& 0& 0& 0& 0 \\
    0& 1& 0& 1& 0& 0& 0& 0 \\
    0& 0& 1& 0& 1& 0& 0& 0 \\
    0& 0& 0& 1& 0& 1& 0& 0 \\
    0& 0& 0& 0& 1& 0& 1& 0 \\
    0& 0& 0& 0& 0& 1& 0& 1 \\
    1& 0& 0& 0& 0& 0& 1& 0 \\
  \end{bmatrix},
  \label{eq:L3}
\end{equation}
where we drop the $-2$ down the main diagonal. In the context of Hamiltonian evolution, the main diagonal contributes a global time-dependent phase and a shift of the eigenvalues by a constant while leaving the eigenvectors unchanged.
Note that the full Laplacian operator is $\frac{1}{a^2}L$.

The results can be generalized to the multidimensional case since contributions in different dimensions are independent. For example, in two dimensions with $A_x$ qubits in the $x$ direction, $A_y$ qubits in the $y$ direction, and independent of the position encoding, we have
\begin{align} \label{eq:LapMultDim}
        & L^{(A_x,A_y)}_{0\ldots (A_x-1), A_x \ldots (A_x+A_y-1)} \nonumber \\
        = & L^{(A_x)}_{0\ldots (A_x-1)} + L^{(A_y)}_{A_x \ldots (A_x+A_y-1)}.
\end{align}
The inexpensive addition of multiple dimensions means that the Laplacian, with Boltzmann statistics, scales linearly with the number of particles.  In general the encoding for a $D$-dimensional lattice is given by layers of one-dimensional codes.  An explicit example of 2 dimensions with 4 lattice positions in $x$ and 8 in $y$ (32 sites in all), encoded in binary with the leading two qubits for $x$ and last three qubits for $y$ is
\begin{equation} \nonumber
\scalemath{0.9}{
  \begin{matrix}
    \begin{matrix} & y\\ x & \end{matrix} & $0$& $1$& $2$& $3$&$4$& $5$& $6$& $7$ \\
     $0$ & $00000$ & $00001$ & $00010$ & $00011$ & $00100$ & $00101$& $00110$& $00111$\\
     $1$ & $01000$ & $01001$ & $01010$ & $01011$ & $01100$ & $01101$& $01110$& $01111$\\
     $2$ & $10000$ & $10001$ & $10010$ & $10011$ & $10100$ & $10101$& $10110$& $10111$\\
     $3$ & $11000$ & $11001$ & $11010$ & $11011$ & $11100$ & $11101$& $11110$& $11111$\\
  \end{matrix}}
  \label{eq:Lap2DExample}
\end{equation}
A two-dimensional example on a larger lattice is implemented and analyzed in Sec.~\ref{sec:sim_quartic}.

In the following sections, we first present the mapping in binary encoding in Sec~\ref{sec:Laplacian:binary} which requires the full Pauli basis, the derivation of the BRGC Laplacian in Sec.~\ref{sec:Laplacian:Gray}, which maps to the $XZ$ model, and the H2GC Laplacian in Sec.~\ref{sec:Laplacian:Dist2Gray}, which maps to the transverse Ising model.

\subsection{The Laplacian matrix in the binary encoding}
\label{sec:Laplacian:binary}

Let $L^{(A, \textrm{bin})}$ be the Laplacian matrix of $2^A$ lattice points with periodic boundary condition in one dimension.
When $A = 3$, for example, $L^{(3, \textrm{bin})}$ is given by Eq. \ref{eq:L3}.
We define the operator
\begin{equation}
  C^{(A)}_{0\dots A-1} =\prod_{i=0}^{A-1} \sigma^+_i + \prod_{i=0}^{A-1}\sigma^-_i. \nonumber
\end{equation}
Then one obtains a recursive formula for $L^{(A,\text{bin})}$:
\begin{equation}
  \begin{split}
  &L^{(A, \textrm{bin})}_{0\dots A-1} = L^{(A-1, \textrm{bin})}_{0\dots A-2}-C^{(A-1)}_{0\dots A-2} + \sx_{A-1} C^{(A-1)}_{0\dots A-2}
  \\
  &= L^{(A{-}1,\textrm{bin})}_{0\ldots A{-}2}  + (\sigma_{A{-}1}^x {-}1)  (\prod\limits_{i=0}^{A{-}2} \sigma_i^x  P_i^0 + \prod\limits_{i=0}^{A{-}2} \sigma_i^x  P_i^1) \hfill
  \end{split}
  \label{eq:L_recursion}
\end{equation}
starting from the two-site Laplacian with periodic boundary conditions
\begin{equation}
    L^{(1)}_{0} = 2\sx_0.
    \label{eq:base_case}
\end{equation}
We emphasize that $L_0^{(1)}$ is the same for all codes and is the starting condition for all recursive formulas presented in this work. As a reminder, the first term in Eq.~\ref{eq:L_recursion} does not include the index $A-1$ and, therefore, there is an implied identity operator on this index, as shown in Eq.~\ref{eq:sz1}. The same convention is applied throughout the paper.

\begin{figure}[ht]
\centering
\includegraphics[scale=0.8 ]{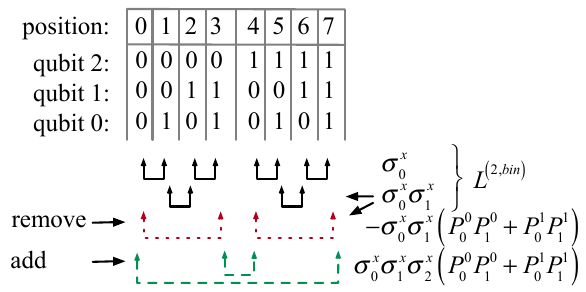}
\caption{The three-qubit binary encoded Laplacian is constructed from the two-qubit Laplacian and two corrections.
The columns under the position index line are the position encoding in qubit states.  For example, position state 4 in the second subblock is encoded as $100$ in qubit states specified below position index 4.
The horseshoe shaped lines with arrows indicate the symmetric contributions between neighboring positions.
The first three row of lines show the contributions inherited from $L^{(2,\textrm{bin})}$, including contributions shown as red (dotted) lines that are excess in the $A=3$ context.  The excess contribution is removed by term $-\sigma_0^x \sigma_1^x \left(P_0^0 P_1^0 + P_0^1 P_1^1\right)$.
The green(dashed) lines indicate the two contributions not provided by $L^{(2,\textrm{bin})}$ that are added by the term to the right.
The corrections are shown next to the red and green lines. }
\label{fig:BinaryProjLap}
\end{figure}

\figref{fig:BinaryProjLap} gives a graphical derivation of  Eq. \ref{eq:L_recursion} for $A=3$.
The sum of projection operator products can be seen to be picking out the ends of the $2^{A-1}$ position subregions.
Then, the $\sigma^x$ operators add the new green dashed contributions in, and subtract the old red dotted contributions out.

The $\sigma_i^x$ part of the correction generates many copies of the product $\sigma_0^x\sigma_0^z=-i\sigma^y_0$, so the Laplacian includes all three Pauli matrices if expanded.
In conjunction with the potential, this mapping uses the entire Pauli basis.
Current quantum annealers can only evolve qubit systems from the transverse-field Hamiltonian to the classical Ising model~\cite{boothby2021architectural}.
While hardware improvements may be developed to handle couplings of other Pauli products, such as $XZ$, in the near future, one expects that it will still handle only Hamiltonians composed of terms with small Pauli support\footnote{The support of an operator is the number of non-identity Pauli matricies.}.

In adiabatic quantum computing, if the target Hamiltonian is classical, \textit{i.e.} diagonal in the computational basis, the reduction of multiqubit to two-qubit interactions is well understood \cite{PhysRevA.78.012320}.
In Eq.~(\ref{eq:L_recursion}), however, the operator is nondiagonal in the computational basis on most of the sites, making the method in Ref.~\cite{PhysRevA.78.012320} not applicable.
In Sec.~\ref{sec:Laplacian:Gray}, we introduce the BRGC encoding of the Laplacian which allows for reduction with $\mathrm{O}(A)$ qubits to a 2-local form.
This is  an important advantage of a BRGC encoding over the binary one.

\subsection{The Laplacian matrix in the binary reflected Gray encoding}
\label{sec:Laplacian:Gray}
As explained in the previous section, the tensor-product decomposition of the Laplacian matrix in the binary encoding has the undesirable $\sy$ terms.
Thus, it is natural to ask whether it is possible to find
a position encoding so that the tensor-product decomposition of the Laplacian is simpler.
We show that the BRGC encoding of position, matching position $x$ to the qubit state specified by the $x^{\textrm{th}}$ member of the BRGC, achieves a dramatic simplification.

An implementable qubit or spin Hamiltonian in current quantum annealers is the sum of transverse fields:
\begin{equation}
  H^x = \sum_{i=0}^{A-1} \sigma^x_i.
\end{equation}
For a system with $A$ qubits, $H^x$ contains $2^{A-1}A$ symmetric couplings between qubit states differing in one bit.
By symmetric, we mean that $H^x$ couples, for example, qubit states $000$ to $001$ and $001$ to $000$.
Each of these symmetric couplings will give rise to two 1s in $H^x$ expressed in the tensor products of the $A$ qubits.
We call the symmetric coupling a connection between the qubit states.
Note that the Laplacian matrix couples consecutive positions.
Thus, if the $A$-body qubit states are ordered so that the consecutive states differ only in the state of one qubit, then the representation of $H^x$ in the ordering contains all the nonzero elements of the Laplacian matrix\footnote{Note that binary encoding does not meet this condition.}.
Gray codes, well-known in signal processing, have this property~\cite{Kautz1958,Tompkins1956,Chinal1973}. We give a short review below.

\subsubsection{Gray encoding}
\label{sec:graycode}
Any Gray encoding (reflection-based or not) of the positions guarantees that neighboring bit strings differ in exactly one bit.
Gray code is an alternate compact binary encoding of integers $0$ to $2^A{-}1$ into $A$ bits.
For example, the standard binary encoding of the numbers from 0 to 3 is 00, 01, 10, 11.
The binary encoding is convenient for arithmetic, but neighboring numbers have varying numbers of bit differences.
For neighborhood operators like the Laplacian we would like to minimize the bit difference between nearby points.
Gray code does this, resulting in neighboring points differing in only one bit in their code.
If we are working in a periodic space, this property is preserved with a one bit difference between the first and last point.
The most common Gray code is the binary reflected Gray code, which can be constructed by induction on $A$.
For the base case, $A=1$, we encode $\{0, 1\}$ as $\{0, 1\}$.
For larger $A$ we concatenate the codes for the $A-1$ case to the reflected (or reversed order) codes for the $A-1$ case.
We then add a most significant bit of 0 to the first half and a 1 to the second half.
For $A=2$, this procedure yields $\{00, 01, 11, 10\}$, and for $A=3$ it yields $\{000, 001, 011, 010, 110, 111, 101, 100\}$.

In general, for an encoding $G$, let the {\it encoding function} $G(n)$ denote the integer that the $n^{\textrm{th}}$ bit string of $G$ represents in binary.
When we have a specific encoding function, we explicitly use its code name to denote it.
For example,
\begin{equation} \nonumber
  \begin{split}
  &\text{BRGC}(0){=}0,\, \text{BRGC}(1){=}1,\,\text{BRGC}(2){=}3,\, \text{BRGC}(3) {=}2\\
  &\text{BRGC}(4){=}6,\, \text{BRGC}(5) {=} 7,\, \text{BRGC}(6){=}5,\, \text{BRGC}(7) {=} 4.
\end{split}
\end{equation}
To transform a matrix $M$ between different encodings, we view the matrix as intrinsically defined with respect to the $A$-body qubit states.
Because the qubit states are ordered differently in different encodings, the matrix transformation is induced by the encoding function $G$.
The encoding function for the binary encoding is the identity map, and therefore $M$ is just $M$.
For the $M$ in encoding $G$, we define
\begin{equation}
  M^{(A,\text{G})}[G(k), G(m)] \equiv M^{(A,\text{bin})}[k,m].
  \label{eq:gray_transform}
\end{equation}
For simplicity, we also denote the $n^{\textrm{th}}$ bit string of $G$ with $G(n)$.

\subsubsection{The recursive formula for the Laplacian matrix}
In the case of the Laplacian, $L^{(\textrm{bin})}$ is tridiagonal, but $L^{(A, \textrm{BRGC})}$ is not.
As explained above, $L^{(A,\text{BRGC})}$ should be closer to $H^x$ than $L^{(A,\text{bin})}$.
For example, when $A = 2$,
\begin{align}
    L^{(2,\textrm{bin})} =
    \kbordermatrix{
        & 0 & 1 & 2 & 3\\
        0 & 0 & 1 & 0 & 1 \\
        1 & 1 & 0 & 1 & 0\\
        2 & 0 & 1 & 0 & 1\\
        3 & 1 & 0 & 1 & 0}
    \xrightarrow[\textrm{to \textrm{BRGC}}]{\text{bin}}
    \kbordermatrix{
        & 0 & 1 & 3 & 2 \\
      0 & 0 & 1 & 1 & 0 \\
      1 & 1 & 0 & 0 & 1\\
      3 & 1 & 0 & 0 & 1\\
      2 & 0 & 1 & 1 & 0
    } \nonumber \\
    = \sum^1_{i=0} \sigma^x_i = H^x
    = L^{(2,\textrm{BRGC})} \nonumber
\end{align}
In the Gray encoding, the $2^A$ connections between adjacent positions in the Laplacian matrix are generated by $H^x$, but there are $2^{A-1}A - 2^A=2^{A-1}(A-2)$ extra connections that must be  suppressed.
We suppress the extra contributions by multiplying by a sum of projection operators on one or more bits.
The projection operators are the simplest if they represent a projection onto a lower dimension subspace of the $2^A$ binary hypercube.

With a BRGC encoding of the positions, the Laplacian has a recursive decomposition that
follows the recursive definition of the BRGC itself.
Because the two sub-blocks are reflections of each other with leading 0 and 1 bits added,
we know precisely the codes at the sub-block boundaries.
The outer codes are all zero except for the leading bit.  The inner codes are zero except for the two leading bits being 01 and 11.   All corrections required to construct the larger $A$-bit Laplacian from the $A-1$ bit Laplacian take place
in the subspace of the 4 codes and are restricted to that subspace by a simple projection operator.

\begin{figure}[ht]
\centering
\includegraphics[scale=0.8 ]{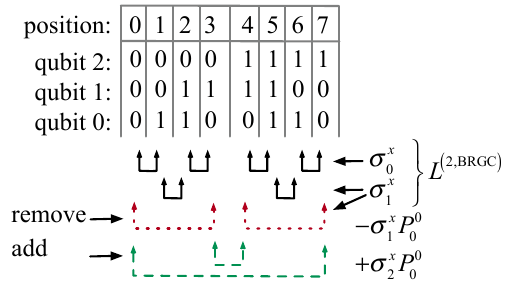}
\caption{The three-qubit BRGC Laplacian is constructed from the two-qubit Laplacian and two corrections.
The columns under the position index line are the position encoding in qubit states.  For example, position state 4 in the second sub-block is encoded as $110$ in qubit states.
The horseshoe-shaped lines with arrows indicate the symmetric contributions between adjacent positions with $\sigma_0^x$ generating the first row and $\sigma_1^x$ the second and third rows.  The third row, drawn as a red (dotted) line, is an excess contribution generated by  $\sigma_1^x$  in the $A=3$ context and is removed by $-\sigma_1^x P_0^0$.   The last row, drawn with green(dashed) lines, shows the missing sub-block end contributions between positions pairs $\{0,7\}$ and $\{3,4\}$, which are added with term $\sigma_2^x P_0^0$.
}
\label{fig:GrayLap}
\end{figure}

The base case is the one-qubit Laplacian of Eq.~(\ref{eq:base_case}).
We emphasize, however, for $A=2$, the neighbor contributions to the Laplacian operator
are
\begin{equation} \label{eq:gray2lapbase}
    L^{(2,\textrm{BRGC})}_{0, 1} = \sigma_0^x + \sigma_1^x,
\end{equation}
which we again recognize as the transverse Hamiltonian acting on qubits $0$ and $1$.
For larger $A$ we have
\begin{equation} \label{eqn:graylap}
    L^{(A,\textrm{BRGC})}_{0\ldots A-1} = L^{(A-1,\textrm{BRGC})}_{0\ldots A-2} +\left(\sigma_{A-1}^x - \sigma_{A-2}^x\right) \prod\limits_{i = 0}^{A - 3} {P_i^0}.
\end{equation}
The reduction to a 2-local form of the projector product is addressed in Sec. \ref{sec:red2local}.
For $L^{(A,\textrm{BRGC})}$,  the decomposition of all projectors result in an additional $\mathrm{O}(A)$ ancillary qubits and 2-local terms.
A graphical derivation of Eq. \ref{eqn:graylap} is given in \figref{fig:GrayLap}.
A more detailed derivation is given in App.  \ref{sec:alt_deriv}.
In this construction, it is clear that since projection operators contain only constants and $\sigma^z$ operators, and that no products of Pauli matrices are taken that act on the same qubits, that each iteration will not introduce either $\sigma^y$ or products of $\sigma^x$.  Since the base case \eqnref{eq:gray2lapbase} has no $\sigma^y$ operators or $\sigma^x$ products, this property is preserved for all $A$.

\figref{fig:GrayLap} shows how the added and subtracted terms correct the 1D Laplacian $L^{(2, \textrm{BRGC})}$ to make $L^{(3, \textrm{BRGC})}$.
All the required corrections take place in the subspace defined by the lower $A-2$ qubits being zero.
The trailing product of projection operators in \eqnref{eqn:graylap} implements the projection onto that subspace.
In a system with $A$ qubits, $L^{(A-1)}$ makes extra $2^{A-1}$  periodic contributions, shown as red dotted lines, which must be removed,
as well as missing contributions connecting the ends of the two $2^{A-1}$ length sub-blocks, shown as dashed green lines, that must be added.

\subsubsection{Reduction to the 2-local form}
\label{sec:red2local}
One can immediately see that the correction terms in each iteration of Eq. \ref{eqn:graylap} are simple, containing none of the $XX$ couplings appearing in $L^{(\text{bin})}$.
In particular, in Eq. \ref{eqn:graylap}, the nondiagonal operator $\sigma^x_i$ appears just once each term.
Thus, one can reduce the shared product of $A-2$ projection operators $P_i^0$ into a single qubit, for which the classical method in Ref.~\cite{PhysRevA.78.012320} applies.
This gives the 2-local Laplacian containing $XZ$ couplings.
It is important to note that the number of ancillary qubits required is a linear function of $A$, preserving the exponential improvement over classical computing.
See App. \ref{sec:ProjectionReduction} for the reduction of $L^{(A,\text{BRGC})}$ to a 2-local Hamiltonian.

\subsection{The Laplacian matrix in the Hamming-distance-2 Gray encoding}
\label{sec:Laplacian:Dist2Gray}

The advantage of a Gray encoding of position is that neighboring positions always differ in a single bit flip,
meaning that $H^x$ automatically generates the neighbor contributions to the Laplacian.
The Hamming distance between two codes is defined to be the number of bit differences between them.
The existence of pairs of nonadjacent positions of Hamming distance 1 requires the use of projection operators to eliminate the associated unwanted contributions.
This motivates a second Gray code sequence that we call a Hamming-distance-2 Gray code sequence.
With H2GC  any two codes in the sequence that are not sequential neighbors are at least Hamming distance 2 from each other.
Finding the longest such sequence is known as the snake-in-a-box problem; optimal sequences are unknown for large $A$.
However, lower-bound constructions show that the sequence length grows exponentially \cite{ABBOTT1991111}, as does the number of omitted codes.
These longest-length H2GCs are quite irregular and would have large and complex penalty Hamiltonian contributions for the excluded codes.

\begin{figure}[h]
\centering
     \includegraphics[width=0.25\textwidth]{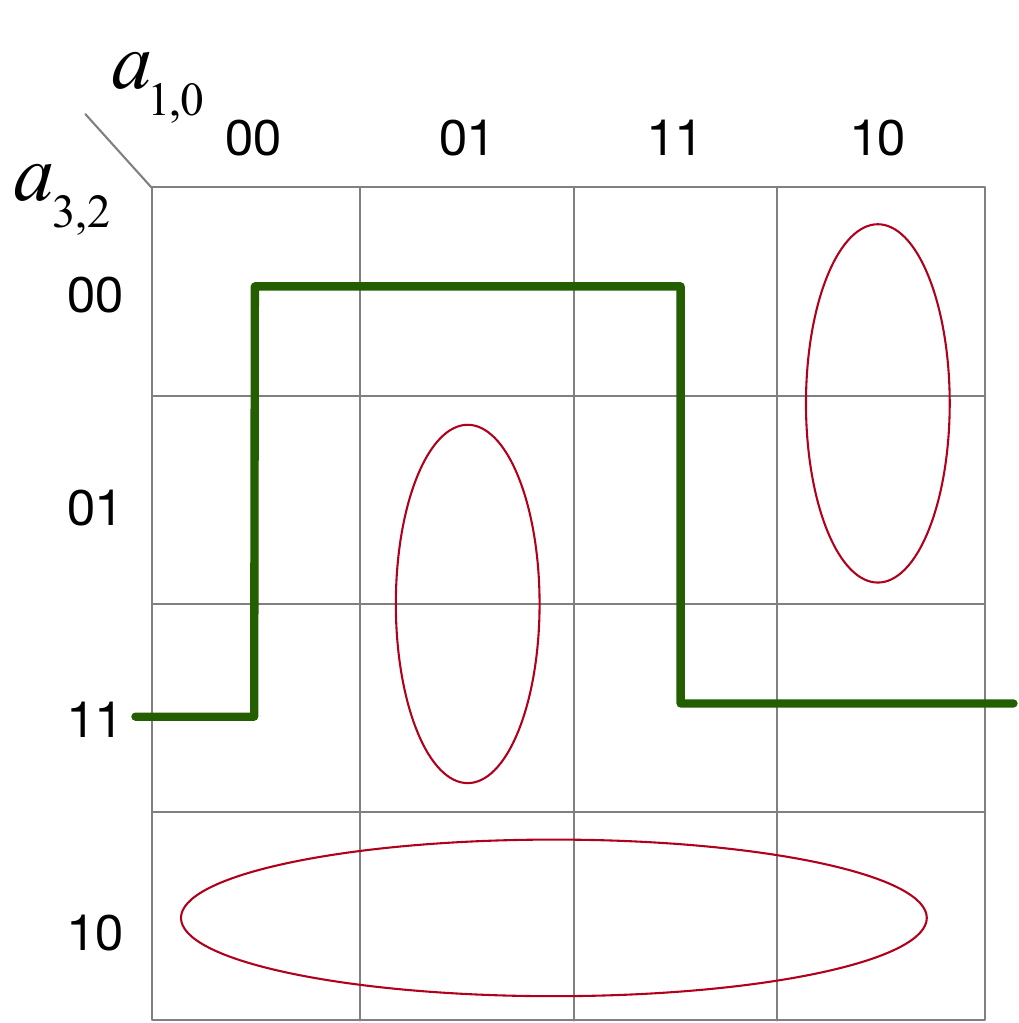}
\caption{The Karnaugh map of the $A=4$ base case for recursive H2GC construction.  The sequence is illustrated with a solid line and passes through eight codes.  The remaining eight unused codes are covered by a 2-local term on the bottom, and a pair of 3-local terms indicated by vertical ovals.}
\label{fig:H2GC_base}
\end{figure}

We now demonstrate a recursive regular construction of an H2GC sequence in which the length grows slightly slower, as $n=2^{A/2+1}$, but with an efficient implementation of the penalty Hamiltonian.
We begin with an $A=4$ construction with sequence length 8 as a base case, illustrated in \figref{fig:H2GC_base} on Karnaugh map.

\begin{figure}[h]
\centering
     \includegraphics[width=0.42\textwidth]{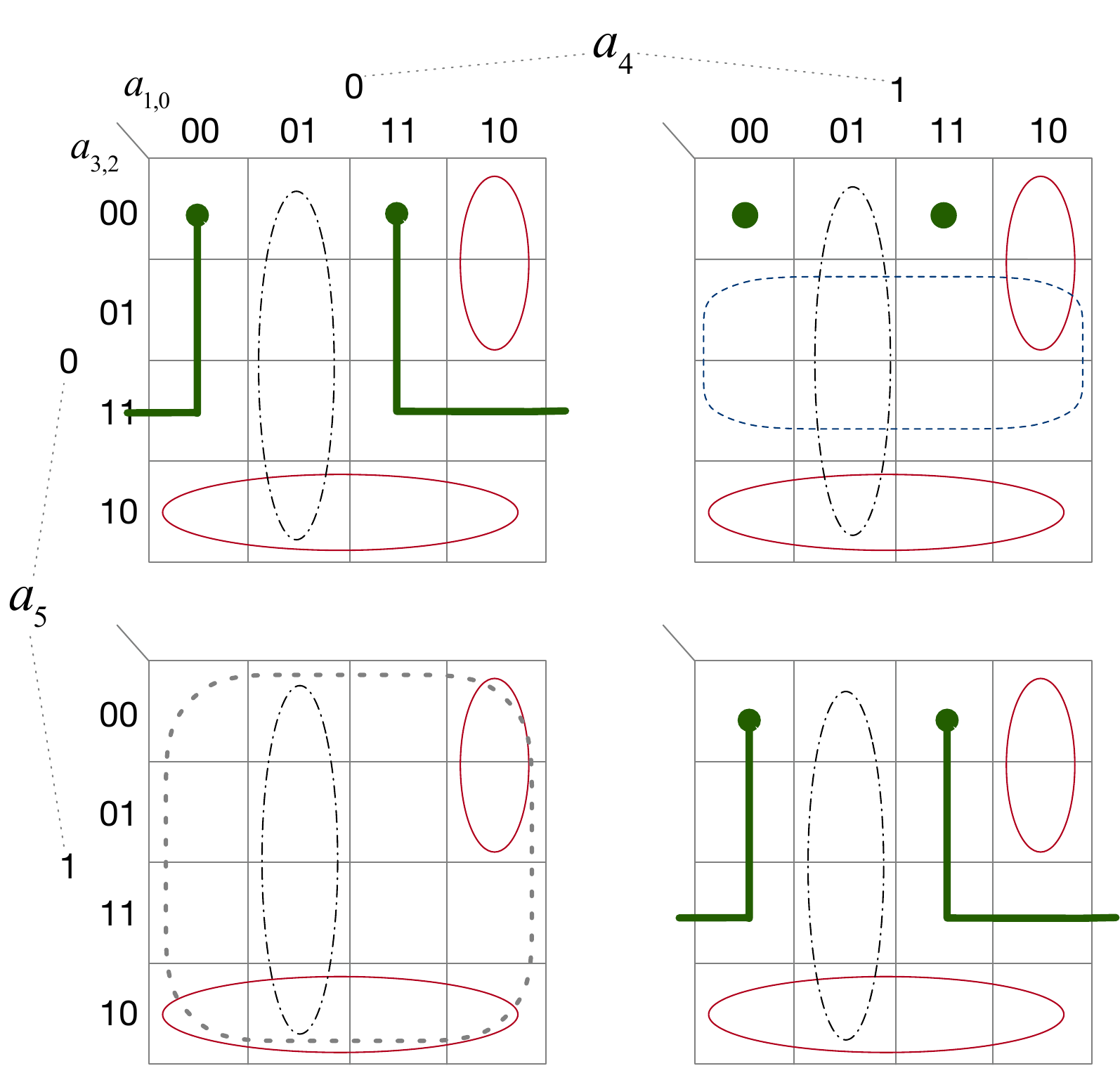}
\caption{The Karnaugh map for the recursive construction of the cycle for $(A+2)$ qubits creates two copies of the $A$ qubit cycle with a gap, isolated by a difference of both new qubits.  Solid ovals are the original penalty cover extended over the two new qubits.   The upper right subspace contains the two codes used to connect the copies into a new cycle.  They are referred to as drill-through codes because they connect the subsequent copies through the isolating layer represented by the upper right subspace.   The additional penalty cover for the upper-right subspace is a set of large subcubes, each covering half the subspace. Code values for $a_{3,2}$ and $a_{1,0}$ should be applied across the entire figure.}
\label{fig:H2GC_R}
\end{figure}
The recursive step builds the $A+2$ qubit sequence from two copies of  the $A$-qubit sequence.  The first step is the removal of a link in the cycle for $A$, exposing two ends that will be used to join the two copies in a larger cycle via codes in the upper-right subspace of \figref{fig:H2GC_R}.
For $A=4$ we choose code $a_{3,2,1,0}=0001$ to remove, as the required penalty term is only 2-local.
For larger $A$ we choose one of the codes in the upper-right subspace previously used to join two subsequences.
A 3-local penalty term suffices to cover the opened code without overlapping the H2GC sequence.
For $A=4$, this is obvious, and for larger $A$, the links between the $A-2$ sequence copies are isolated in a subspace encoded by $a_{A+1,A} = 01$ ($a_{5,4}=01$ in the figure) and a single variable suffices to distinguish the two link codes in the subspace, which can be seen in the upper right $a_{5,4}=01$ subspace of \figref{fig:H2GC_R}.

We then make two copies of the $A$ sequence associating one with a code of $a_{A+1,A} =00$ and the other with a code of $a_{A+1,A} =11$.
 In doing so, all existing penalty terms extend over all four values of the new qubits, and are therefore preserved unchanged for the $A+2$ sequence.

 The penalty cover for the bottom-left subspace, which the sequence does not enter,  is completed by a single 2-local term that depends only on the two new qubits.
 The upper-right subspace has two drill-through sequence members added that complete the new H2GC cycle.
 The penalty cover for the unused part of the upper-right subspace is completed by $(A-3)$ 3-local terms, each including one subspace qubit having the same value for both drill-through codes, with the penalty term carrying the the opposite value for the qubit, and a code of 01 for the new qubits.  It would be $(A-2)$ 3-local terms, but the inherited penalty terms will already cover $1/4$ of the subspace, seen in the bottom row
 of the subspace.  At larger $A$, the pre-existing coverage comes from the lower-left subspace of the $A-2$ sequence.

 The upper-left and lower-right subspaces are themselves distance-2 sequences because they are derived from the known good sequence on $A$ qubits.   All members of the upper left and lower right differ in both of the new qubits.
The drill-through codes in the upper right differ in one bit with the adjacent sequence member in the upper left and lower right subspace, and must differ in an additional bit with any other sequence member in the upper-left or lower-right.
And last, the two drill-through codes differ by two bits from each other by construction.
We conclude therefore that the constructed sequence is a H2GC sequence.

 The construction of the $(A+2)$ sequence, adds $(A-2)$ 3-local terms to the penalty Hamiltonian.   Therefore, the number of 3-local terms grows quadratically with $A$.   A reduction to 2-local  requires at most (some ancillary qubits may be shared) $(A-2)$ additional ancillary qubits.

A key observation is that the Hamiltonian for the system now takes the form
\begin{equation} \nonumber
	\sum\limits_{i=0}^{A-1}{\sigma_i^x} + A(t)\, Q\!\! \!\!\!\! \!\!\sum\limits_{p\in \textrm{penalty terms}} {  \!\!\!\!\!\!\!\!p } +\;  B(t) V.
\end{equation}
With the penalty strength, $Q$, set to $\mathrm{O}(1/a)$, the spurious degrees of freedom decouple from the theory. For the decomposition of the potential, the prohibited basis states are then ``don't cares'', a term from Boolean logic minimization, meaning
that the value of $V(c)$ for the prohibited codes may be adjusted to a value that simplifies the decomposition of $V$.

In this Hamiltonian, all terms are either just $\sigma^x$ or products of $\sigma^z$ (from the expansion of the projection operators), and therefore map to the $k$-local transverse Ising model.   The set of projection operators can be decomposed to a 2-local form using intermediate qubits with penalty terms as is discussed in App.~\ref{sec:ProjectionReduction}, which then further maps the Schr\"odinger equation to a 2-local transverse Ising model.
The simple form of the H2GC Laplacian raises the possibility of applying this technique with existing hardware systems.

One strategy for evolving to the H2GC Laplacian is to first turn on the penalty via $A(t)$ and later to turn on $V$ via $B(t)$. We provide a demonstration of the time evolution of the H2GC Hamiltonian in Sec.~\ref{sec:sim_ho}.

Alternatively, since the ground state of the H2GC Laplacian operator is known, one can imagine initializing a future quantum computer with this \textit{a priori} known (and therefore ``trivial'' in the context of the quantum adiabatic theorem) ground state.
Specifically, at first approximation, the $n$ valid sites the ground-state wave function are given by the normalization coefficient, while the invalid sites are then set to zero such that
\begin{align}
\Psi(t=0, x) = & 1/2^{(A+2)/4} && \forall x \in \{\textrm{H2GC}(i)\},\nonumber \\
\Psi(t=0, x) =& 0 && \forall x \notin \{\textrm{H2GC}(i)\}\nonumber,
\end{align}
where $\{i \in \mathbb{Z} | 0 \leq i < 2^{A/2+1}\}$. Corrections to this ideal ground state are polynomially suppressed by the strength of the penalty coefficient.
The exact form of higher-order correction terms will depend on the penalty cover as shown in Fig.~\ref{fig:H2GC_R} and is beyond the scope of the current work.
The time evolution would then depend only on the intermediate ground-state gap induced by $V$ as its contribution is turned on, and importantly, be protected by the IR cutoff of the system, which we discuss in more detail in Sec.~\ref{sec:simulations}.

\section{POTENTIAL DECOMPOSITION}
\label{sec:potential_decomposition}
We demonstrate the method of mapping any local, real potentials sampled at $N=2^A$ lattice sites in $D$ dimensions.
The potential matrix is diagonal and spanned by the product of $A$-body $\sz_i$ interactions yielding the complete Walsh basis in encoding $G$:
\begin{equation}
  \begin{split}
  W^G_n &=  (\sz)^{\text{bit}(G(n),A-1)} \otimes \cdots  \otimes (\sz)^{\text{bit}(G(n),0)}\\
  &= \bigotimes_{i=A-1}^{0} (\sz)^{\textrm{bit}(G(n), i)}
  \end{split}
\end{equation}
where $G(n)$ is the $n$th bit string in $G$.
Note that $\text{bit}(s,A-1)$ is the most significant, i.e. the leftmost, bit of $s$, and $\text{bit}(s,0)$ is the rightmost bit.
Independent of the encoding, the Walsh functions are bijectively mapped to the set of $k$-local Ising models with $k \leq A$, and, indeed, any local potential can be represented in this fashion.
This conclusion can be reached by recognizing the Walsh basis as the digitized version of the Fourier series.
While the focus of this work are quantum adiabatic compuations, we would like to emphasize that this bijection reduces the problem of constructing the minimal depth quantum circuit for a given error tolerance, for an arbitrary diagonal unitary operator $e^{i f(\hat{x})}$, to that of finding the minimal length Walsh-series approximation of the exponent $f(x)$~\cite{Welch2014}.
Details of the Walsh basis and their relationship to the Ising model and Fourier series are provided in App.~\ref{sec:math_review}.

Typically, given a local potential $V(x)$ in the continuum, its discrete version $V^{\text{bin}}[m]$ is an array obtained through sampling the continuum at successive lattice spacings $a$:
\begin{align}
  V^{\textrm{bin}}[m] = V^{\textrm{cont.}}(m a).
\end{align}
In order to correctly evaluate the Schr\"odinger equation however, one must encode the position of the discretized potential in the same encoding $G$ as the Laplacian operator $L^{(A, \textrm{$G$})}$,
\begin{align}
 V^{\textrm{G}}[ G(m) ] = V^{\textrm{bin}}[m],
\end{align}
For example, if we want to solve the Schr\"odinger equation in BRGC with $L^{(A, \textrm{BRGC})}$, then a two-qubit potential will take the form
\begin{equation}
  \begin{split}
   V^\text{BRGC}[0]  =  &V^\text{BRGC}[\text{BRGC}(0)] = V^\text{bin}[0]\\
   V^\text{BRGC}[1]  =  &V^\text{BRGC}[\text{BRGC}(1)] = V^\text{bin}[1]\\
   V^\text{BRGC}[2]  =  &V^\text{BRGC}[\text{BRGC}(3)] = V^\text{bin}[3]\\
   V^\text{BRGC}[3]  =  &V^\text{BRGC}[\text{BRGC}(2)] = V^\text{bin}[2]\\
  \end{split}
\nonumber
\end{equation}
because the positions $[0, 1, 2, 3]$ are encoded in BRGC as $[00, 01, 11, 10]$ and reinterpreted as binary numbers to $[0, 1, 3, 2]$ in the same way the Laplacian is encoded.

After being encoded, the potential can then be expanded in $W^G_n$.
Typically, the inner product with each basis element is taken, resulting in a series expansion.
However, in practice, this is computationally expensive as, for $N=2^A$, $\mathrm{O}(N^2)$ operations are required.
To speed up the decomposition, the fast Wash-Hadamard transform (FWHT) is employed to reduce the complexity to $\mathrm{O}(N \log(N))$ operations. This transform expands any real potential in $W^G_n$ to a given order.

As a consequence of the various codes available, the FWHT is also representation dependent. However, the chosen representation in this case is immaterial and yields only a remapping of the Walsh functions. In particular, given a system of $A$ qubits, one obtains the same set of basis operators.
For example, if one works in the binary representation, what is labeled $W^{\textrm{bin}}_2$ will simply be bijectively remapped to $W^{\textrm{BRGC}}_3$ as discussed in App.~\ref{sec:math_review}, while the resulting $k$-local Ising model stays unchanged. In this paper, we use the sequency ordered transformation $\textrm{FWHT}^{\textrm{seq}_A}$~\cite{Pratt1969,Manz1972} because the subscript label in $W^{\textrm{seq}_A}_n$ can be interpreted as the sequency{\footnote{The sequency of a Walsh function is the number of positive zero-crossings of that function.} of the basis function and is therefore the choice that mimics the Fourier-series mode expansion.

In summary, the steps of mapping the potential to the qubit or spin Hamiltonian are: 1) discretize the potential to a given lattice, 2) map the potential array to the same code as the Laplacian, 3) decompose the mapped potential using FWHT, 4) map the resulting series expansion to the $k$-local Ising model.

\subsection{Potential coarse graining}

While the FWHT reduces the complexity of decomposition, the cost still scales exponentially with respect to the number of qubits.
 To further reduce the setup cost, we opt to employ coarse-graining methods.
If the features of the potential are on a scale that is much larger than the lattice spacing, then one expects a low-mode expansion to be a sufficient representation of the potential.
As a result, given a coarse-graining scale $a^{\textrm{CG}} \geq a$, the complexity of the FWHT becomes $\mathrm{O}(N^{\textrm{CG}}\log(N^{\textrm{CG}}))$  where $N^{\textrm{CG}} = (L/a^{\textrm{CG}})^D$ in $D$ dimensions can be exponentially smaller than the original lattice. Such a strategy allows one to decouple the setup cost from the lattice size for a suitable potential.

In this work we explore two coarse-graining strategies: averaging and decimation.
In both cases, we define the coarse-grained lattice, $N^{\textrm{CG}}=2^{A^{\textrm{CG}}}$ where  $A^{\textrm{CG}}<A$.

In averaging, we block average the potential between a given interval. This approach has the benefit of obtaining exactly the same coefficients as in the complete expansion with the high-sequency modes $A^{\textrm{CG}}<r\leq A$ set to zero, serving as a low-pass filter.
Therefore for suitable potentials, averaging introduces only a series truncation error that is well behaved, in the sense that the coefficients of higher-sequency contributions are at least polynomially suppressed.
If we have the functional form of the potential, one can analytically compute the indefinite integral and construct averages for the result, and would be computationally cheap to carry out.
If the potential does not have an analytic form, one would require sampling at all grid locations making the computational complexity of averaging $\mathrm{O}(2^{A})$, and can become prohibitively expensive.

One approach to coarse-grain potentials without analytic forms is to sample only $N^{\textrm{CG}}$ equidistant points through decimation.
Unlike averaging however, decimation introduces an uncontrolled error in the values of the coefficients in addition to the series truncation error.
Multiple coarse-graining scales will need to be studied to numerically demonstrate that decimation is under control.
Nevertheless, the computational cost of decimation can be made negligibly small for any potential.

\subsection{Example: $S$-wave deuteron potential}
As an illustration, in Fig.~\ref{fig:pot_walsh_normal}a, we plot both the potential and its low mode expansions from both strategies.
We construct a simple $S$-channel smooth hard core plus well nucleon-nucleon potential that roughly mimics the form of the well-known Argonne $v_{18}$ potential \cite{AV18_PhysRevC.51.38}.
The height of the hard core and the depth of the well are tuned to reproduce the deuteron binding energy in infinite volume.
The potential has the functional form
\begin{align}
 V_{NN}(r) = E_{\textrm{core}} e^{-(r/{R_\textrm{core}})^4} - E_{\textrm{well}} e^{-(r/{R_\textrm{well}})^4},
 \label{eq:nn_potential}
\end{align}
where $E$ and $R$ are free parameters tuned to experimental data parameterizing the height and radius of the core ($r\lesssim 0.3$~fm) and well ($r\gtrsim 0.3$~fm).
Details of the free parameters are given in Fig. \ref{fig:pot_walsh_normal}a.

\begin{figure}[ht]
 \begin{tabular}{c}
 \includegraphics[scale=0.5]{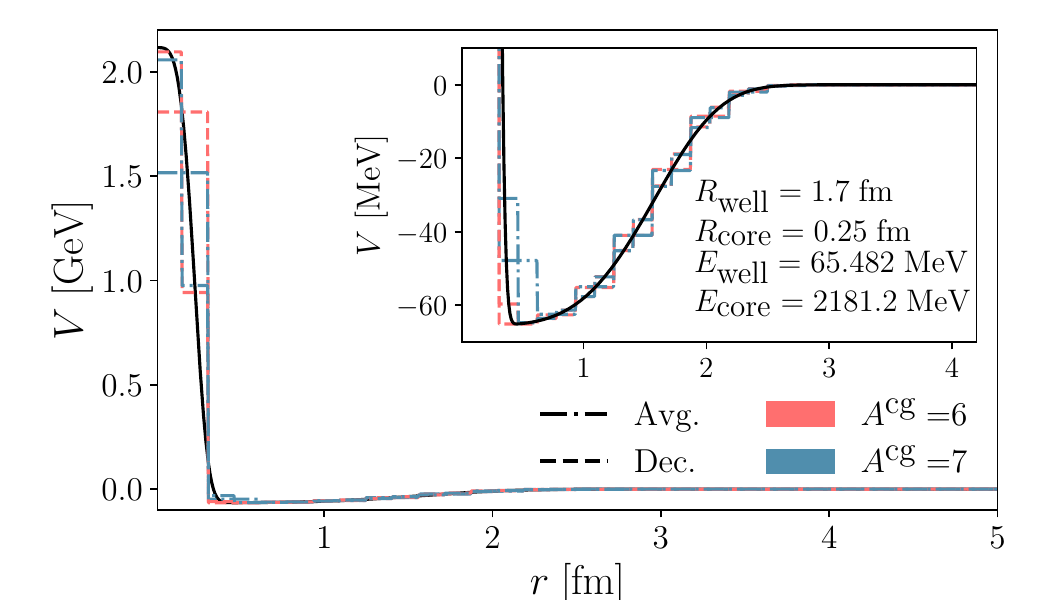}\\
 \textbf{a}\\
 \includegraphics[scale=0.5]{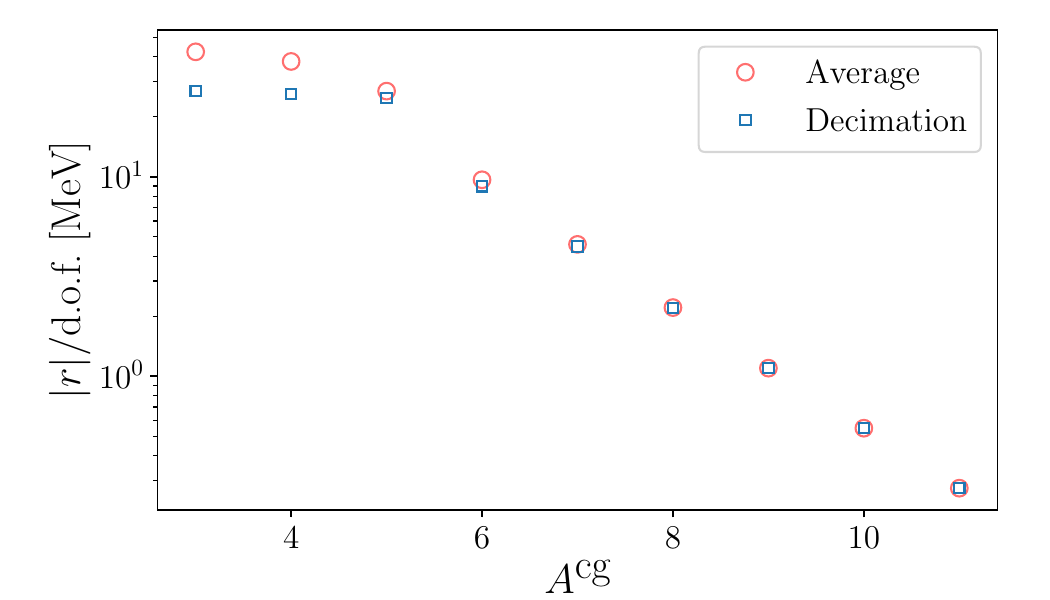}\\
 \textbf{b}
 \end{tabular}
 \caption{
 \textbf{a)} The $S$-wave potential \eqnref{eq:nn_potential}, with parameters in the inset tuned to reproduce the deuteron binding energy (solid black line), and the respective discretized potentials on coarse-grained lattices. The inset graph enlarges into the bottom of the well near $r=0.5$ fm to show how the discretization approximates the potential there.
 \textbf{b)} The $L_1$-normed error per degree of freedom (DOF) of the coarse-grained potential as a function of the number of qubits used to represent the coarse-grained lattice. As $A^{\textrm{CG}}$ increases, the $L_1$-normed difference from the continuum potential exponentially decreases.}
 \label{fig:pot_walsh_normal}
\end{figure}

As demonstrated, the Walsh expansion is a very effective representation in both approaches. The difference between averaging and decimation decreases rapidly as $A^{\textrm{CG}}$ increases. The difference from the continuum potential is summarized in Fig.~\ref{fig:pot_walsh_normal}b, where the $L_1$-normed error per degree-of-freedom is shown to decrease exponentially given a linear increase in $A^{\textrm{CG}}$.

The potential is encoded in the BRGC representation for this example.
Once encoded, it is then expanded by the aforementioned $\text{FWHT}^{\textrm{seq}_A}$.
As an illustration, we provide the Walsh expansion for the averaging scheme depicted in Fig.~\ref{fig:pot_walsh_normal} for $A^{\textrm{CG}}=4$,
\begin{align}
 V_{NN} &\xrightarrow[]{\text{discretize}} V^{\textrm{bin}}_{NN} \xrightarrow[]{\text{BRGC}} V^{\textrm{BRGC}}_{NN}\nonumber \\
 = & 129 \sum_{i=0}^{3} W^{\textrm{seq}_4}_i + 129 \sum_{i=4}^{7} W^{\textrm{seq}_4}_i \nonumber \\
 &+ 135 \sum_{i=8}^{10} W^{\textrm{seq}_4}_i + 135 \sum_{i=11}^{13} W^{\textrm{seq}_4}_i, \nonumber
\end{align}
where, for brevity, we round to integer values in the decomposition.
Since we have the functional form of the potential given by, Eq.~(\ref{eq:nn_potential}), the averaging can be performed analytically.

The resulting potential decomposition to the $k$-local Ising model is
\begin{align}
  H^{\textrm{BRGC}}_{V} = & 129 \left( I + \sigma^z_3 + \sigma^z_3 \sigma^z_2 +  \sigma^z_2\right)\nonumber \\
  +& 129 \left( \sigma^z_2 \sigma^z_1 + \sigma^z_3 \sigma^z_2 \sigma^z_1  + \sigma^z_3 \sigma^z_1 + \sigma^z_1\right)  \nonumber \\
  +& 135 \left( \sigma^z_1 \sigma^z_0 + \sigma^z_3 \sigma^z_1 \sigma^z_0 + \sigma^z_3 \sigma^z_2 \sigma^z_1 \sigma^z_0 + \sigma^z_2 \sigma^z_1 \sigma^z_0\right) \nonumber \\
  +& 135  \left( \sigma^z_2 \sigma^z_0 + \sigma^z_3 \sigma^z_2 \sigma^z_0 + \sigma^z_3 \sigma^z_0  + \sigma^z_0\right), \nonumber
\end{align}
which can be obtained by inspection from the bijective map between the Walsh functions and the $k$-local Ising model.
The $k$-local terms can be decomposed to a binary tree of ancillary qubits constructed with 2-local terms~\cite{PhysRevA.78.012320, Welch2014} and in App. \ref{sec:ProjectionReduction}.
We comment that the deuteron potential requires a relatively large basis to describe because the 2~GeV hard core is $\delta$-function-like, and poses a challenge for series-expansion methods.

\section{ADIABATIC QUANTUM-COMPUTING SIMULATIONS}
\label{sec:simulations}
\begin{figure*}[t]
 \begin{tabular}{cc}
 \includegraphics[scale=0.5]{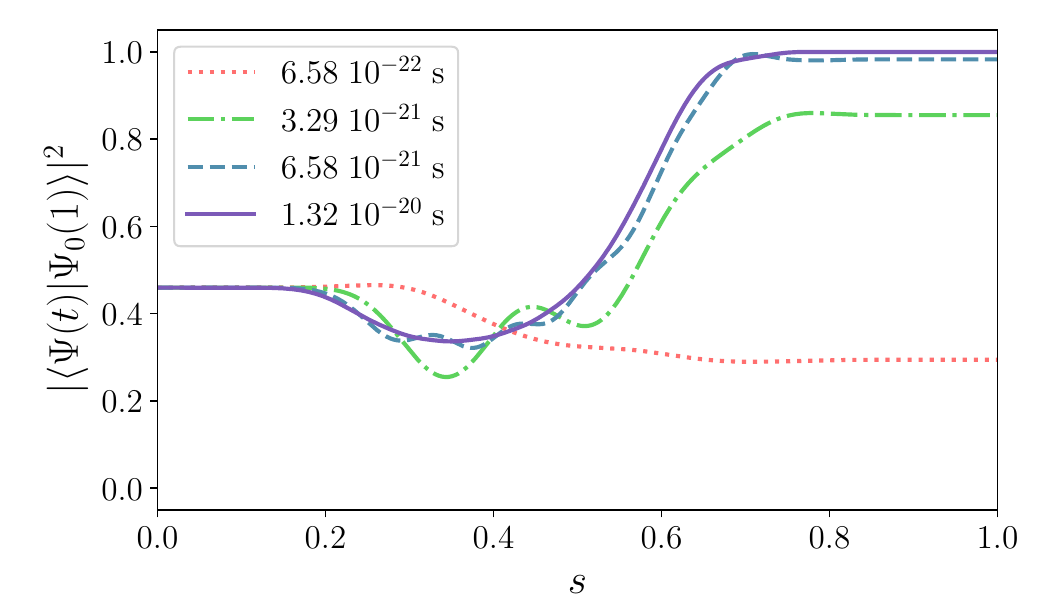} &
  \includegraphics[scale=0.5]{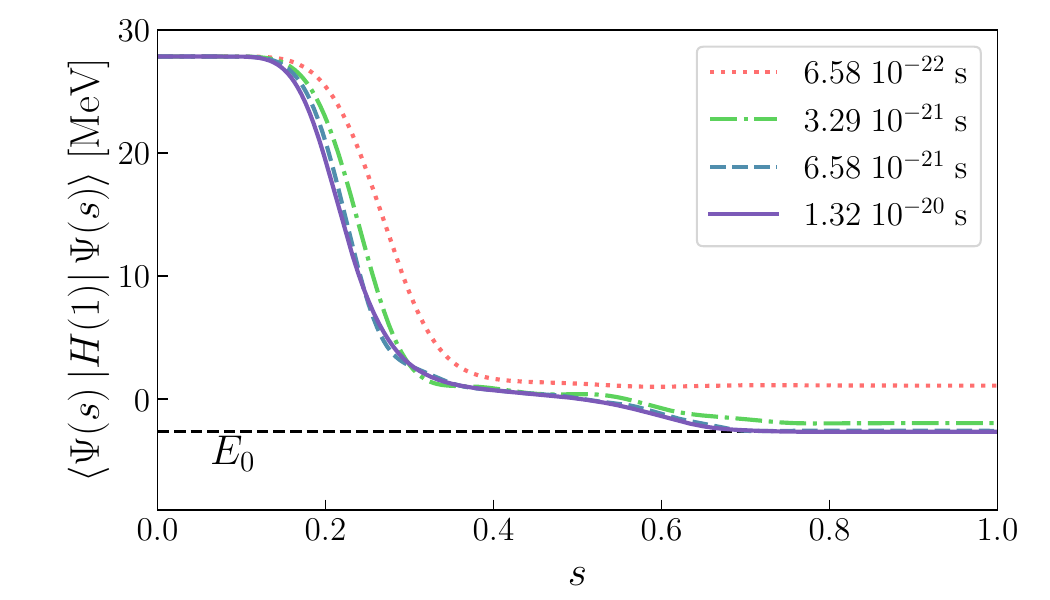}\\
 \textbf{a} & \textbf{b}
 \end{tabular}
 \caption{Adiabatic evolution from a free particle to the interacting system with the $S$-wave potential of Sec.~\ref{sec:potential_decomposition}.
 \textbf{a)} Overlap with the true ground state and expectation of the interacting Hamiltonian as a function of total evolution time.
 \textbf{b)} The ground-state energy as a function of total evolution time. The dashed black line marks the physical deuteron binding energy of -2.2 MeV.
 }
  \label{fig:anneal_deuteron}
\end{figure*}

Adiabatic quantum computation (AQC) solves for the ground state of a complex Hamiltonian by starting from the known ground state of a trivial Hamiltonian and adiabatically evolving the initial Hamiltonian to the final target~\cite{Kadowaki1998, 2000quant.ph..1106F, RevModPhys.80.1061}.  AQC is an alternative paradigm for realizing universal quantum computation.
Quantum annealing hardware is the closest to an implementation of AQC.
It solves problems where the initial Hamiltonian is the transverse field, and the final Hamiltonian is restricted to be a 2-local Ising model.
One goal of this work is to tailor our algorithm to be implemented with as few extensions to existing hardware as possible in the hope that new generations of hardware will incorporate them.
In particular, the application of BRGC eliminates the necessity of $\sy$. It requires only the addition of a 2-local $XZ$ coupling. The application of H2GC allows simulations to proceed via the transverse Ising model and in principle, could be implemented today if the transverse field were allowed to persist throughout the evolution.

In this section we simulate the following time-dependent Hamiltonian
 \begin{align}
    H(s) = & L + B(s) V, \label{eq:aqc_h}\\
    \Psi(0) = & \left(\left|\uparrow\right> + \left|\downarrow\right>\right)^{\otimes A},
    \label{eq:ht}
\end{align}
where $L$ is the Laplacian defined in Sec.~\ref{sec:laplacian_construction}.
Here, $a=L/2^A$ is the lattice spacing and $V$ is the potential from Sec.~\ref{sec:potential_decomposition}.

The time dependence comes from $B(s)$, whose argument s is a dimensionless coefficient with normalized evolution time $s=t/T$ with total evolution time $T$ such that $s\in [0, 1]$.
The initial wave function $\Psi(0)$ is an equal superposition state in any encoding of the Laplacian.
In particular, the ground state of $L$ for binary encoding and BRGC is exactly the same as the transverse Hamiltonian $H^x=\sum_i \sx_i$, which is seen to be  the zero-frequency plane-wave solution.
It follows that adiabatic evolution to $H(1)$ prepares the qubits into the ground state of a given quantum system, which becomes the starting point for a time-dependent Schr\"odinger simulation.

In~\cite{Kadowaki1998} the authors compare quantum and statistical annealing for the transverse Ising model using three annealing schedules; linear, square root, and the logarithmic form.  They found that the logarithmic annealing schedule keeps the wave function closest to the instantaneous ground state (with the largest overlap). In this work, we chose the schedule based on recent developments in understanding adiabatic schedules~\cite{Albash2018,Dong2020}. Following Ref.~\cite{Dong2020}, we employ a schedule with vanishing gradients at the boundary,
\begin{equation}
 B(s)=\frac{\int_0^{s}\exp \left( \frac{-1}{s^\prime(1-s^\prime)}\right)ds^\prime}{\int_0^{1}\exp \left( \frac{-1}{s^\prime(1-s^\prime)}\right)ds^\prime}
 \label{eq:anneal_schedule}
\end{equation}
for all simulations presented in this work. Additional optimizations to the schedule warrant further investigation~\cite{PhysRevX.8.031016, 2020RPPh...83e4401H, doi:10.7566/JPSJ.89.044001}, but are beyond the scope of this work.

The total evolution time $T$ can be roughly estimated from the IR cutoff of the theory given by the box size. In particular, the energy gap between the ground state and first excited state of the free field equation with periodic boundary conditions is
\begin{equation}
 \delta E = \frac{\left(2 \pi \right)^2}{2mL^2} \gg 1/T
 \label{eq:adiabatic_bound_napkin}
\end{equation}
where $m$ is the (reduced) particle mass, and $L$ is the length of the finite box. In the examples below, we find setting $T$ to be approximately 2 orders of magnitude longer than $1/\delta E$ is sufficient to evolve the system adiabatically. We note that there exist proofs of rigorous bounds for the quantum adiabatic theorem~\cite{Jansen_2007, Albash2018}, but when applied to examples discussed later in this section, the rigorous bounds overestimate the required time by several orders of magnitude when compared to both Eq.~(\ref{eq:adiabatic_bound_napkin}) and observations from the corresponding numerical simulations.
Additional investigation of tighter theoretical adiabatic bounds for Hamiltonian simulation is important but beyond this work's scope.

In the other extreme, the UV cutoff is regularized by the lattice spacing and given by $1/a$.
The critical role of the UV cutoff in the application of the H2GC Laplacian is discussed in Sec.~\ref{sec:sim_ho}.

All simulations have been performed with \texttt{QbSim}, a quantum bit simulator.  \texttt{QbSim}  performs real or imaginary time simulation of qubit systems where the Hamiltonian is expressed as
\begin{equation}
H(t) = \sum\limits_{i}   B_i(t) H_i .
\end{equation}
The $B_i(t)$ functions are scalar weight functions implementing time dependence.  The $H_i$  components are expressed as sums of products of Pauli matrix operators and composites like projection operators $P_i^v$.
A Python integration is used to configure the simulation and access the system's evolving state as time is advanced.
A higher-order Dyson series expansion with automatic step size control generates the state evolution.
Calculations take place in a fully parallel way. With GPU acceleration, runtimes are reasonable for 20+ qubit systems.
We intend to write a separate document describing \texttt{QbSim}, and make it available for broader use.

In the sections that follow, we continue the discussion of the $S$-wave deuteron potential mapped using BRGC as a time-independent application, followed by an example of a time-dependent calculation in two dimensions for a quartic potential in BRGC, and conclude with a simple harmonic oscillator mapped to H2GC.

\subsection{Example: $S$-wave nucleon potential with BRGC}
\label{sec:sim_deuteron}

We simulate quantum adiabatic evolution for the potential discussed in Sec.~\ref{sec:potential_decomposition}. The grid size is chosen to be $N=2^7$ based on the results displayed in Fig.~\ref{fig:pot_walsh_normal}. The $S$-wave potential is fitted to reproduce the deuteron binding energy with a reduced mass of $469.14$ MeV,  and vanishes at a distance $r \approx 5$ fm. Because the system is weakly bound it extends much further in radius. The box size is set to $L = 20$ fm to properly represent the interacting system's ground state.

In Fig.~\ref{fig:anneal_deuteron} we show the evolution of the system as a function of total evolution times ranging from $(1~\textrm{MeV})^{-1}$ to $(50~\textrm{keV})^{-1}$. Given a lattice box size of 20~fm, the lowest nonzero momentum state is approximately a 4-MeV excitation above the ground state. We observe that at a total evolution time of $(100~\textrm{keV})^{-1}$ recovers the ground state at the end of the evolution with 98\% probability, and at $(50~\textrm{keV})^{-1}$ the probability increases to 99.97\%. These values are roughly 2 orders of magnitude longer than the estimate of Eq.~(\ref{eq:adiabatic_bound_napkin}).

\begin{figure}[t]
 \includegraphics[scale=0.5]{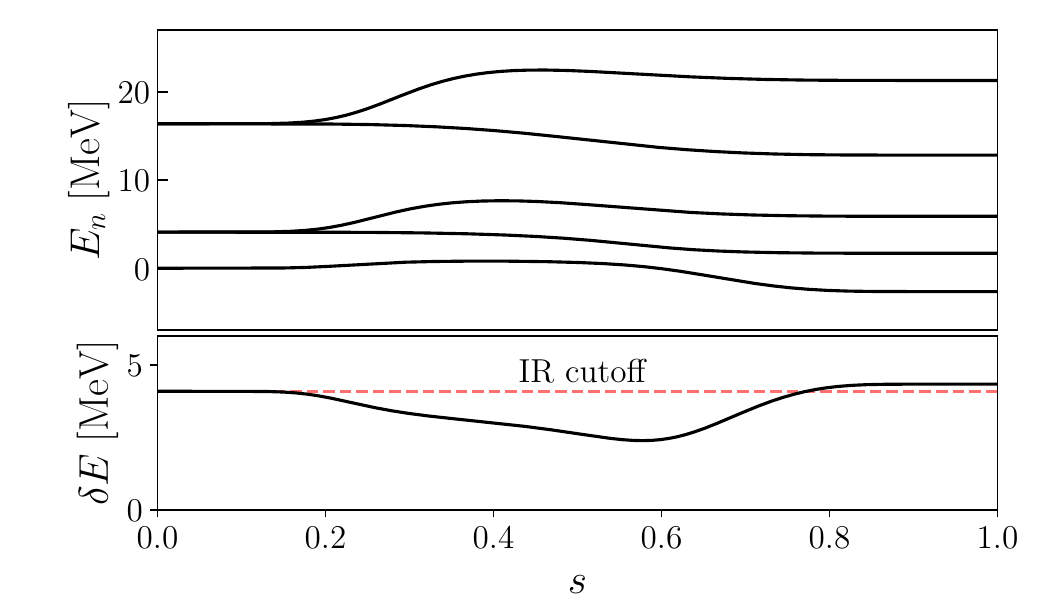}
 \caption{(Top) The instantaneous spectrum during the adiabatic evolution for the first five states. At $s=0$, the free-field Laplacian has a two-fold degeneracy for periodic boundary conditions. The $NN$ potential lifts the degeneracy afterwards. (Bottom) The energy gap between the first excited state and ground state as a function of evolution time $s$. The dashed red line is calculated from Eq.~(\ref{eq:adiabatic_bound_napkin}), yielding the infrared cutoff.}
 \label{fig:deuteron_spectrum}
\end{figure}

While changes in the schedule will affect the result, we observe numerical evidence for physical systems that the IR cutoff of the theory sets the scale for adiabatic evolution. As a result, for a physical system, the total evolution time required for adiabatic state preparation is expected to scale polynomially with the box size, while exhibiting constant scaling with respect to the lattice discretization, which governs the ultraviolet cutoff. This claim is further supported by studying the instantaneous energy spectrum during the evolution as shown in Fig.~\ref{fig:deuteron_spectrum}. We observe throughout the entire evolution that the ground-state to the first excited-state energy gap remains of the same scale as the IR cutoff, only subject to small changes even when the system is undergoing the nontrivial change of introducing a 2-GeV hard-core potential. The fundamental reason why the energy gap is so well protected, even against significant changes in the potential, is that the kinetic energy is quantized within a finite box.
This is a significantly different situation than the typical quantum annealing application, in which the transverse field is progressively switched off during the evolution.

A classical determination of the ground state is more challenging than one might expect.
Because of the large difference in scale between the hard-core height and the binding energy, numerical differential equation solvers are unstable, requiring extra precision and care to find the ground states.
A more straightforward technique is to pick a large discrete basis such as more than O(100) states in a harmonic oscillator basis or a similar number of points in a discrete position basis.
One then takes matrix elements in that basis and diagonalizes.
The large basis is required to simultaneously resolve the spatially tiny hard core and represent the wave function at the long-range associated with the small binding energy.
The runtime of partial diagonalization with techniques like the Lanczos algorithm is a function of $N$, the number of basis states, and the number $c$ of matrix-vector product iterations required, taking $\mathrm{O}(cN^3)$ for a dense matrix.   In contrast, with the position encoding here, $A=7$ qubits yield $N=128$ basis states.
\begin{figure}[ht]
 \includegraphics[scale=0.5]{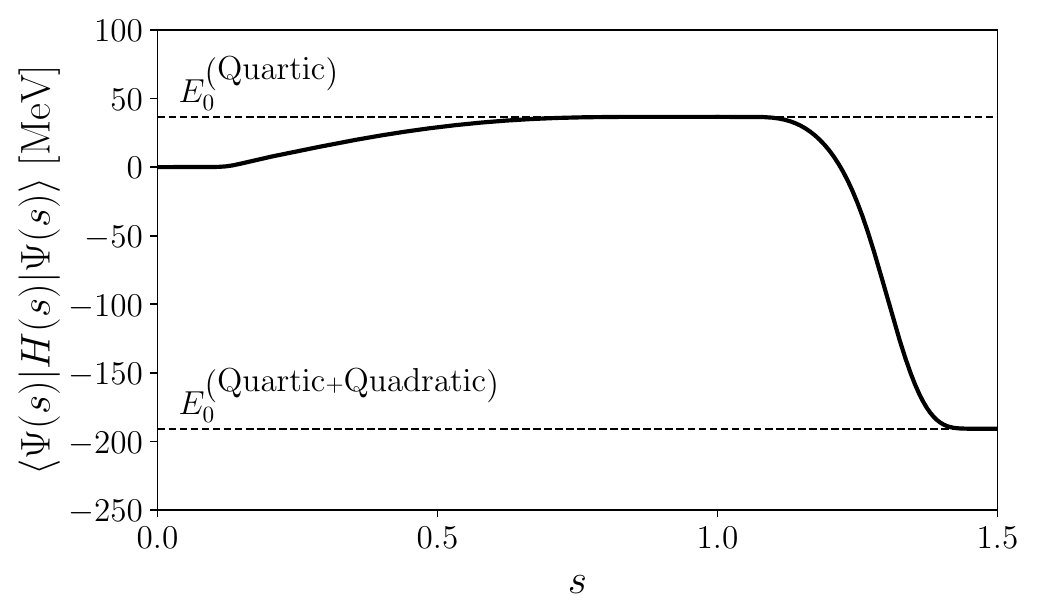}
 \textbf{a}
 \includegraphics[scale=0.5]{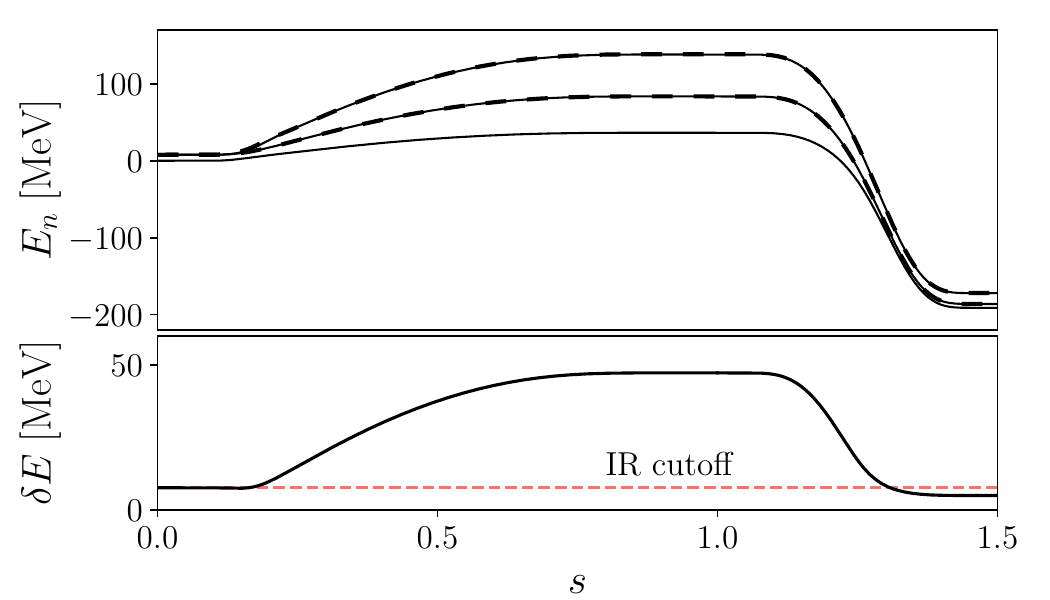}
\textbf{b}
 \caption{ \label{fig:quartic_energy}
\textbf{a)} Instantaneous energy of the time-evolved wave function as function of evolution time $s$. The dashed line labeled by $E_0^{(\textrm{quartic})}$ shows where the ground-state energy of the system with the quartic potential lies.  The dashed line labeled by $E_0^{(\textrm{quartic}+\textrm{quadratic})}$ shows where the ground-state energy lies with the addition of the quadratic potential. \textbf{b)} (Top) The energy spectrum of the first five eigenstates of $H(s)$, and (bottom) energy gap between the ground state and the first excited state as functions of evolution time. The dashed red line is calculated from Eq.~(\ref{eq:adiabatic_bound_napkin}), yielding the infrared cutoff.}
\end{figure}

\subsection{Example: two-dimensional quartic potential with BRGC}
\label{sec:sim_quartic}

In the section we demonstrate the ability to evaluate potentials beyond one dimension.
The example performs adiabatic evolutions in two stages.
Starting from the free particle Hamiltonian, we first evolve the system into a quartic potential as an example of initial state preparation, followed by the introduction of an additional quadratic potential,
\begin{equation}
 \begin{split}
  V^{(\text{quartic})}(x,y)=&V_4(x^2+y^2)^2,\\
  V^{(\text{quadratic})}(x,y)=&-V_2(x^2+y^2),\\
 \end{split}
\end{equation}
where, $V_4 = 10$~MeV and $V_2 = 100$~MeV. As a consequence of turning on the quadratic potential, the ground-state wavefunction which was originally centered around the origin deforms into a ring.

The effective mass is set to $m = 1$~GeV in a box that is $L = 10$~fm long in each direction.
We opt for $N=2^6$ lattice points per dimension (for a total of 12 qubits) yielding a lattice spacing of $a\sim 0.15$~fm resulting in a UV cutoff of approximately $1.3$~GeV. Correspondingly, the IR cutoff is approximately 8~MeV given the box size and particle mass. As a result, we set the evolution time of going from free field to the quartic potential as $T_1=(50~\textrm{keV})^{-1}$, while $T_2=(100~\textrm{keV})^{-1}$ is used as the evolution time for ramping up the quadratic potential.
For purposes of separating the two parts of the evolution, we normalize the evolution time of the first half to $s_1 \in [0,1]$, which governs the initial state preparation of the interacting system with a quartic potential. In the second part of the simulation, $s_2 \in [1, 1.5]$ the quadratic potential is turned on gradually.

\begin{figure}[ht]
 \includegraphics[scale=0.5]{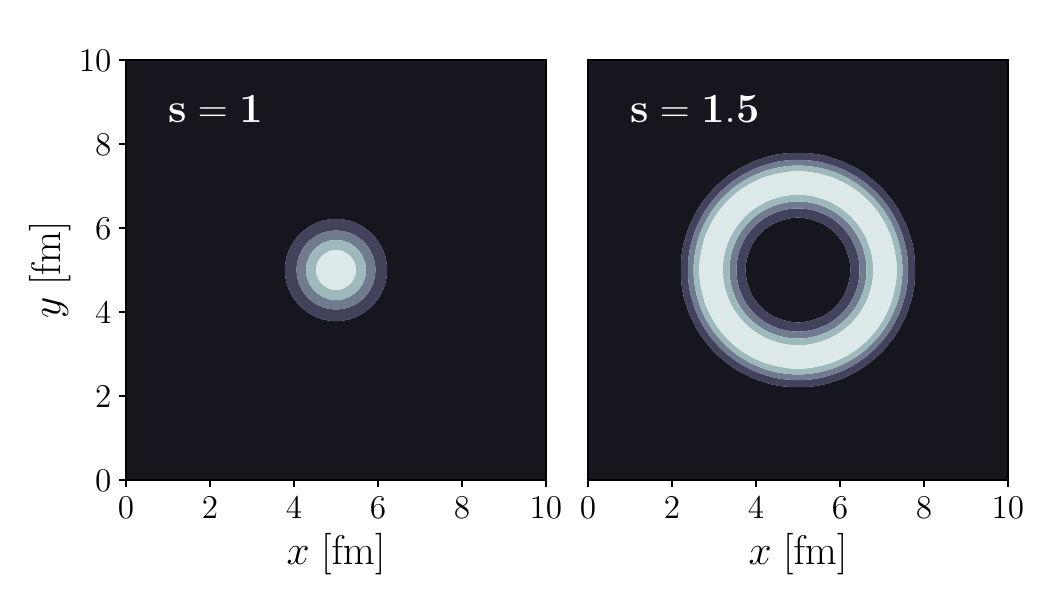}
 \caption{The probability density from the instantaneous wave function at $s=1$ of the quartic potential, and $s=1.5$ with the addition of a quadratic potential.}
 \label{fig:quartic_wf}
\end{figure}

In Fig.~{\ref{fig:quartic_energy}}a, we show the time-dependent energy of the system. Due to the long evolution time, we observe that the system reaches the correct ground state for the quartic potential at $s=1$. The system then proceeds with the addition of a quadratic potential and reaches its new ground-state energy.
Fig.~\ref{fig:quartic_energy}b shows the evolution of the low-lying spectrum of the system.
Similar to the deuteron example in Sec.~\ref{sec:sim_deuteron}, we observe that the minimum energy gap stays well protected by the IR cutoff.
Additionally, in Fig.~\ref{fig:quartic_wf} we show snapshots of the wave function at $s=1$ and $s=1.5$, illustrating the phase transition effects.
The wave function can, in principle, be obtained through repeated measurements of the qubits at the end of the evolution.

\subsection{Example: harmonic oscillator with H2GC}
\label{sec:sim_ho}

In this section, we perform a calculation using the H2GC Laplacian. The final target potential is that of the simple harmonic oscillator
\begin{equation}
V(x) = \frac{1}{2}m x^2.
\end{equation}
In this example, we work in dimensionless units for simplicity, setting the particle mass $m$ to 10 to confine low-lying states to the box and avoid finite volume effects.
We work in a symmetric box ranging from $L=[-1, 1]$.
With $A=8$, H2GC has 32 valid codes, yielding a lattice spacing of $a=2/32=0.0625$.
As a result, the IR cutoff is $\frac{(2\pi)^2}{2mL^2}=0.4934$, and the UV cutoff is 16.
Because the Laplacian matrix is multiplied by $\frac{1}{Ma^2}$ in the Schr\"odinger equation, we use the scale $U \equiv \frac{1}{Ma^2}$ to measure the penalty strength $Q$ needed to suppress the invalid codes in H2GC.

\begin{figure}[ht]
 \begin{tabular}{cc}
 \includegraphics[scale=0.5]{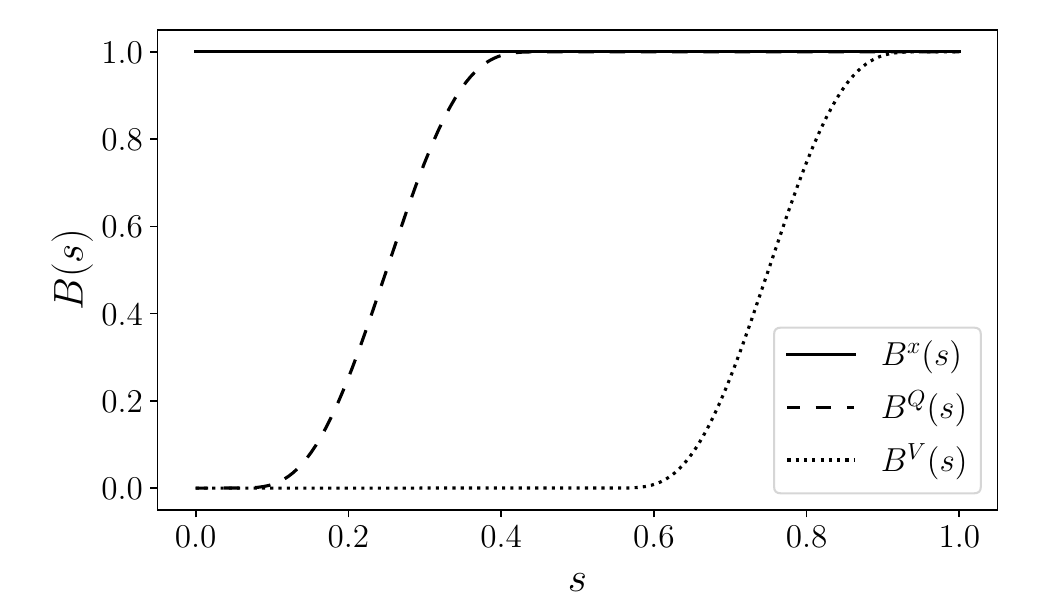}\\
 \textbf{a}\\
  \includegraphics[scale=0.5]{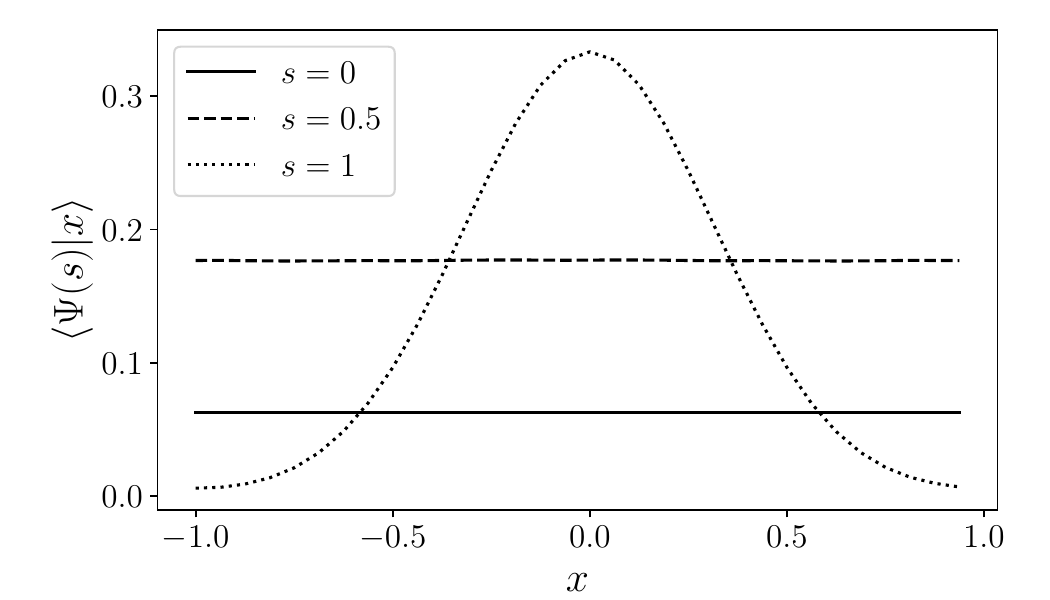}\\
 \textbf{b}
 \end{tabular}
 \caption{ \label{fig:ho_sch_wf}
 \textbf{a)} The time-dependent coefficient $B^x(s)$ for the transverse Hamiltonian $H^x$, $B^Q(s)$ for the penalty Hamiltonian $Q\sum_{c} P(c)$, and $B^V(s)$ for the potential. \textbf{b)} The ground-state eigenvector of $H(s)$ for when (solid line) $s=0$ and the system is governed by $H^x$, (dashed line) $s =  0.5$ when the system is the H2GC Laplacian, and (dotted line) $s=1$ where the harmonic oscillator potential is present.}
\end{figure}

Our time-evolution strategy keeps the transverse-field constant during evolution using the schedule function defined in Eq.~(\ref{eq:anneal_schedule}) to introduce the penalty Hamiltonian, followed by a second delayed schedule to introduce the harmonic oscillator potential. Fig.~\ref{fig:ho_sch_wf}a shows the time-dependent weight $B(s)$ for the three different contributions to the Hamiltonian. The total time required for adiabatic evolution can be roughly estimated by considering the dynamics of two stages: 1) transverse Hamiltonian to H2GC Laplacian and 2) H2GC Laplacian to the harmonic oscillator system.

Given a system of qubits in the ground state of the Laplacian operator, the evolution time required to turn on the harmonic oscillator potential follows the reasoning from previous sections and is some multiple (\textrm{e.g.} $100\times$) the IR scale, which in this example still holds. However, the physics governing the adiabatic evolution to the H2GC Laplacian from the transverse Hamiltonian is dominated by the UV cutoff, and must therefore scale as a function of $U$. This is because the penalty Hamiltonian coefficient needs to be many orders of magnitude above the UV scale to preserve the Laplacian's eigenspectrum. As a result, the penalty Hamiltonian significantly ramps up as a function of $s$ and requires a commensurate amount of evolution time. In simulations, we set the evolution time of the first stage to equal the penalty Hamiltonian coefficient. Applying this logic, the time complexity of adiabatically evolving from the transverse Hamiltonian to the H2GC Laplacian scales exponentially poorly with the number of qubits, since the lattice spacing approaches the continuum exponentially quickly
due to the exponential growth of the length of the H2GC code.
Nevertheless, the H2GC Laplacian provides a way to implement the Schr\"odinger equation on hardware very similar to systems available today, with the additional requirement to retain the contribution of the transverse field throughout the whole evolution.

For a better understanding of what is happening to the wave function, we provide the ground-state eigenvectors obtained from direct diagonalization in Fig.~\ref{fig:ho_sch_wf}b. We observe that when $H(s)$ is the transverse Hamiltonian, the ground state is the properly normalized superposition state. With eight qubits, the normalization factor is $\sqrt{1/2^8}=0.0625$, as indicated by the solid black line.
When the penalty Hamiltonian is fully engaged, the H2GC Laplacian will be in its zero-energy ground state. When properly normalized over 32 positions, we see a constant wave function at $\sqrt{1/32}=0.177$ as shown by the dashed line.
After introducing the harmonic oscillator potential, the wave function becomes the expected Gaussian, as demonstrated by the dotted line.

One can further infer the dynamics of the system by studying the time-dependent spectrum of the system shown in Fig.~\ref{fig:ho_spectrum}a.
In a system of eight qubits, we observe that the system exhibits an eightfold degeneracy in the first excited state, as is expected from the transverse Hamiltonian.
We plot the ninth (dashed odd state) excited state to confirm there are no additional degeneracies. When the penalty Hamiltonian is introduced, the eightfold degeneracy evolves into the expected tower of twofold degeneracies for the Laplacian operator in a periodic box.

Fig.~\ref{fig:ho_spectrum}b further demonstrates why the  prohibited codes must be cleanly gapped from the rest of the system.
In this plot, we show the time-dependent energy gap between the ground state and first excited state. We expect that the free-field Laplacian has a gap given by the IR cutoff, while deviations from the red line arise only from interactions with the potential.
After increasing the penalty coefficient to $100U$, the H2GC Laplacian starts to reproduce the expected gap within 1\% (Fig.~\ref{fig:ho_spectrum}b).
 \begin{figure}[ht]
  \begin{tabular}{cc}
 \includegraphics[scale=0.5]{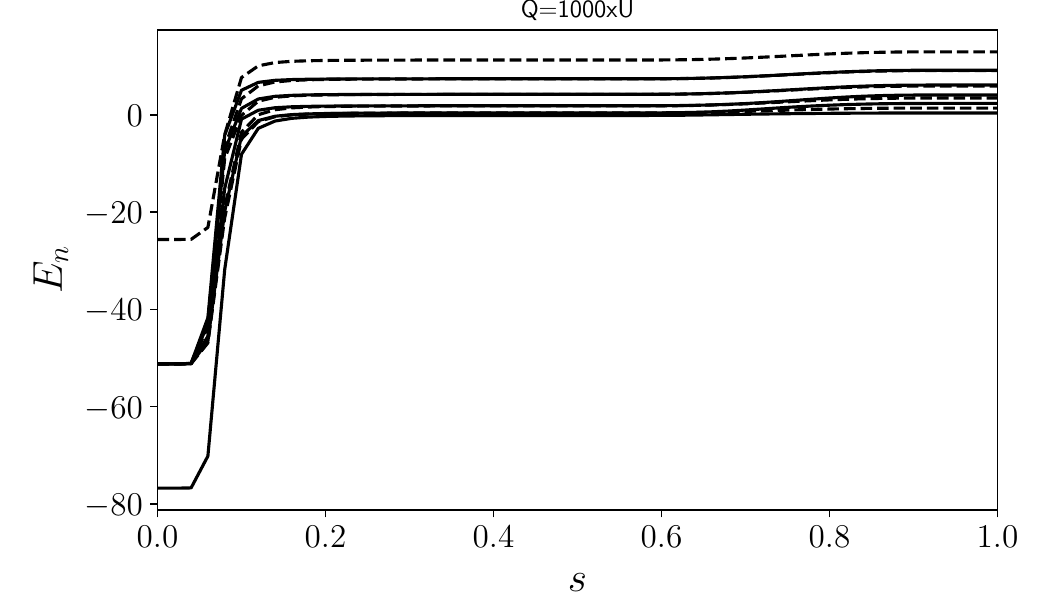}\\
 \textbf{a}\\
 \includegraphics[scale=0.5]{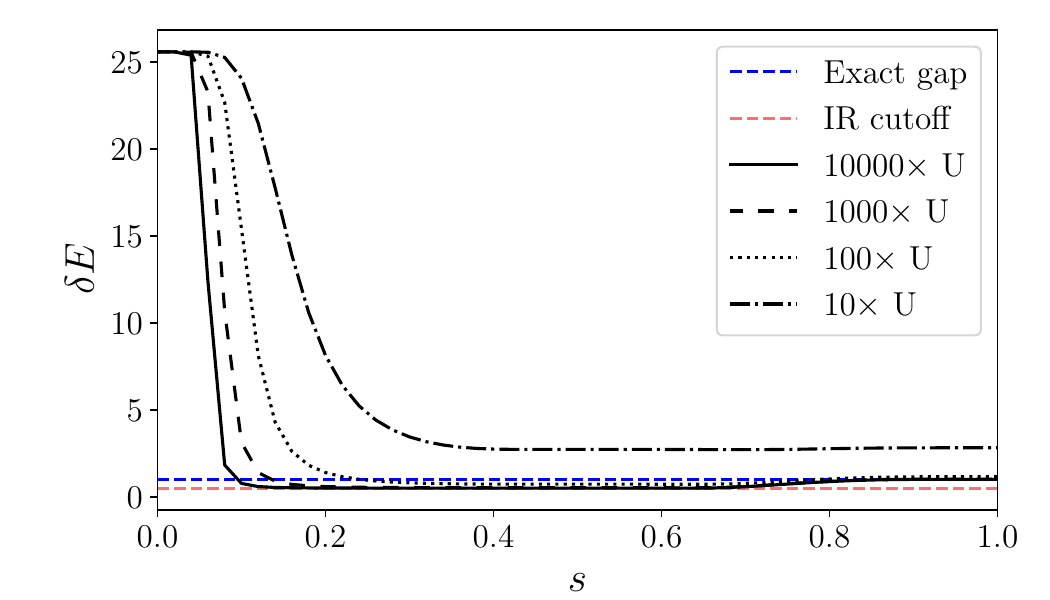}\\
 \textbf{b}
 \end{tabular}
 \caption{  \label{fig:ho_spectrum}
\textbf{a)} The time-dependent energy spectrum of $H(s)$ of the first ten states for a system of eight qubits. The solid lines denote even-numbered states (0, 2, 4, 6, 8), and dashed lines label odd-numbered states (1, 3, 5, 7, 9). \textbf{b)} The time-dependent energy gap between the first excited state and ground state as a function of evolution time $s$ and strength of the penalty coefficient $Q$. The dotted, dashed, and solid lines denote progressively stronger penalty coefficients, set relative to the $U$ scale. The dashed red line is calculated from Eq.~(\ref{eq:adiabatic_bound_napkin}), yielding the infrared cutoff.}
\end{figure}

\subsection{Example: He atom with two electrons}
\label{sec:sim_he}

For a final demonstration we include an example with two independent particles.
We work in units of electronvolts (eV) and nanometers (nm).
We model two electrons, distinguished by spin, around a helium nucleus at the
origin in a periodic three-dimensional volume $b=0.128~nm$ on each side.  The volume size is approximately 4 times the Bohr radius $r_b = 0.031~nm$
found from an effective central charge of approximately $1.69$ due to partial shielding by the other electron.
Each electron spatial direction is BRGC encoded with three qubits for a total of 18 qubits (262144 basis states) and the
the multiparticle Laplacian is implemented following \eqnref{eq:LapMultDim}, summing the Laplacian on the first nine with that of the last nine qubits.
A bare Coulomb potential would yield an infinite sum over periodic images, so
a Yukawa potential (in natural units)
\begin{equation}
V(r) = Z_1 Z_2 \alpha \frac{e^{-r/b}}{r+r_E}
\label{eq:yukawa}
\end{equation}
is used for the interactions between the electrons and between the nucleus and electrons.
The parameters $Z_1$ and $Z_2$ are the charges for particle 1 and 2 respectively, where $Z=-1$ for the electrons and $Z=+2$ for the helium nucleus.
The parameter $\alpha\sim 1/137$ is the fine-structure constant, and $r_E=1.6~fm$ is the helium-nucleus charge radius.
Contributions from neighboring periodic images are included but suppressed by the exponential in the Yukawa potential.    

We first use Lanczos diagonalization to solve for the ground state, yielding a total binding energy of
$74.5$ eV versus the experimental value of approximately $79$ eV.    Given the small number of qubits, the eigenvalue is sensitive to the
volume size as well as the inclusion of an exponential factor in the potential and a close match is not expected.    
We then evolve with AQC from the initial transverse state with $H_0 = T$  to $H=T+V$.

\begin{figure}[ht]
 \includegraphics[scale=0.5]{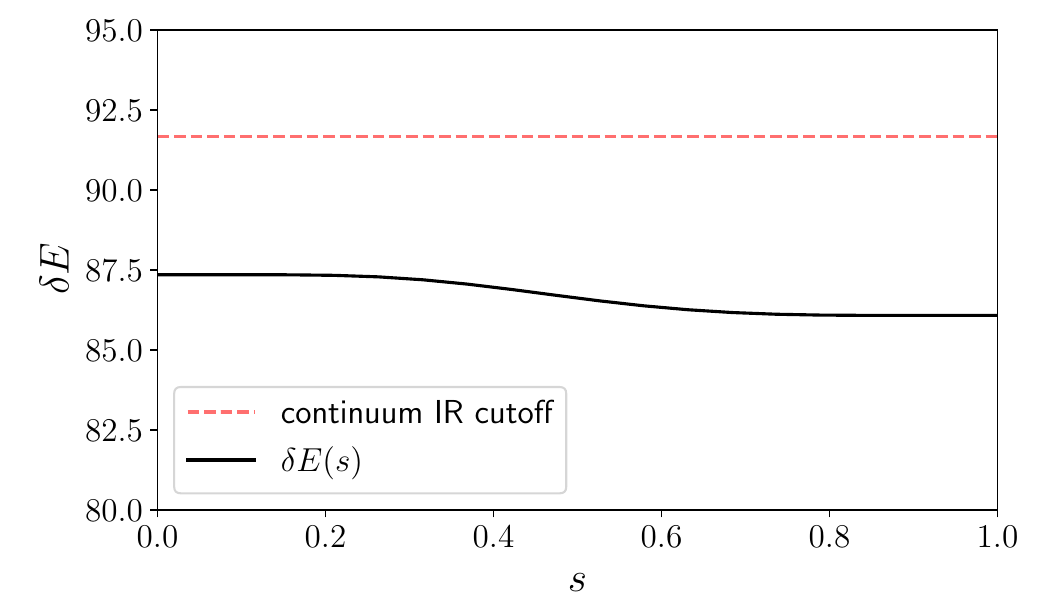}\\
 \caption{  \label{fig:2body-spectrum}
The time-dependent energy gap for $H(s)$ given by Eq.~(\ref{eq:aqc_h}) (black solid line), where the two-particle Yukawa potential in three dimensions (Eq.~\ref{eq:yukawa}) is used. The corresponding energy gap of the free-field system given by Eq.~(\ref{eq:adiabatic_bound_napkin}) (dotted red line) differs from the gap at $H(0)$ with a correction of $\mathrm{O}(a^2)$ at leading order, and is visible due to the relatively coarse lattice spacing used in this example.}
\end{figure}

Fig.~\ref{fig:2body-spectrum} shows the time-dependent energy gap between the ground state and first excited state. The dotted red line is the IR cutoff predicted by Eq.~(\ref{eq:adiabatic_bound_napkin}) given by a 0.128~fm box. Due to the coarse discretization used in this example, the IR cutoff predicted in the continuum limit differs (inconsequentially for our purposes) by approximately 5\% from the gap of the discretized Laplacian. More importantly, we observe that the time-dependent energy gap is again, protected by the IR cutoff and therefore is expected to retain polynomial time complexity with respect to increasing system size.

\begin{figure}[ht]
 \includegraphics[scale=0.5]{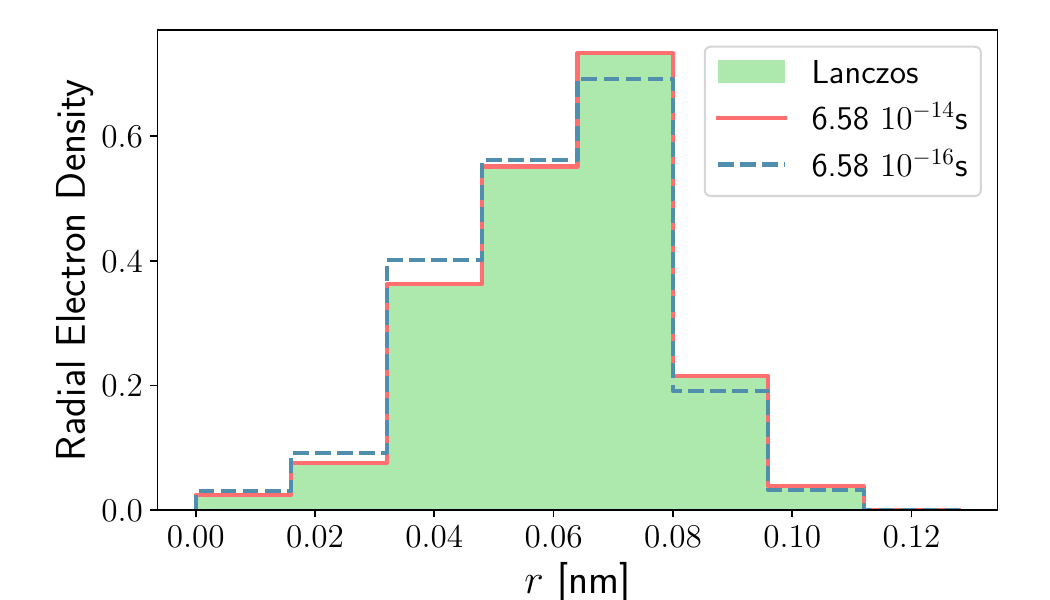}\\
 \caption{  \label{fig:he_dens}
The radial electron density obtained with AQC for $T\sim 1/\delta E$ (dotted blue) and $T\sim 100/\delta E$ (solid red), compared to the density obtained by diagonalization (green histogram) for the He atom, where $\delta E$ is the energy gap between the ground-state and first excited-states of the free-field solution.}
\end{figure}

In \figref{fig:he_dens} we compare the density determined by AQC to that determined from diagonalization.
We observe that when a total evolution time of $100\times$ the IR cutoff is used for AQC, the electron density reproduces the result from exact diagonalization within a fraction of a percent.
We highlight here the observation that the adiabatic bound of $\mathrm{O}(100/\delta E)$ is consistent with all other examples presented in this work.

\section{SUMMARY AND CONCLUSION}
\label{sec:summary}
The Schr\"odinger equation remains one of the foundational blocks of our understanding of quantum systems. One common method of solving this equation is by discretization
in a selected basis and it has found wide applications in classical computations.
We introduced the concept of encoding in the association of positions with $A$-body qubit states in the computational basis.
Such an association provides an exponential improvement in space complexity, and encoding further reduces the Hamiltonian complexity.
With the limitations of current adiabatic quantum computers in mind, we demonstrate the power of Gray encoding in simplifying the spin Hamiltonian representation.
With a BRGC encoding of positions, we mapped the Schr\"odinger equation with a real, local potential to the $XZ$ model, and through H2GC we further simplified it to a transverse Ising model while maintaining an exponential improvement in space complexity and a quadratic count of 3-local penalty terms for unused codes.    These advantages also apply to the implementation of ladder operators.
More specifically, with H2GC, the generic Schr\"odinger equation with a local potential discretized on $2^{A/2+1}$ points is equivalent to a transverse Ising model on $A$ qubits and a quadratic in $A$ count  of 3-local penalty terms.

In both BRGC and H2GC cases, we employed the FWHT to efficiently encode the potential as an Ising Hamiltonian and showed that coarse-graining techniques
could further reduce the computational cost of encoding.

Through numerical simulations, we discovered that the system's adiabatic evolution is stable due to the infrared cutoff associated with finite volume.
By borrowing techniques successfully used in lattice QCD computations with finite range interactions, \textit{e.g.} Luscher's method~\cite{Luscher1986}, for extracting infinite volume results from finite volume ones,
we can envisage performing quantum calculations in finite volume and benefiting from the polynomial time scaling associated with finite volume for computing observables.

For all codes, evolution from a free field to an interacting system exhibits polynomial time complexity with volume and constant scaling with respect to lattice discretization.
For H2GC, if the evolution begins with the transverse Hamiltonian followed by the introduction of penalties to keep the low-lying spectrum of the Laplacian intact, the time evolution will initially be sensitive to the ultraviolet scale. This sensitivity will give rise to polynomial time complexity with lattice discretization.

\newpage

%========================================================================================
\section{ACKNOWLEDGEMENTS}
%========================================================================================
We thank Dong An and Alessandro Roggero for useful discussions and suggestions.

Lawrence Berkeley National Laboratory (LBNL) is operated by The Regents of
the University of California (UC) for the U.S. Department of Energy (DOE) under
Federal Prime Agreement DE-AC02-05CH11231.
This material is based upon work supported by the U.S. Department of Energy,
Office of Science, Office of Nuclear Physics,
Quantum Horizons: QIS Research and Innovation for Nuclear Science
under Award Number FWP-NQISCCAWL (CCC, KSM, YW).
E.R. acknowledges the NSF N3AS Physics Frontier Center, NSF Grant No. PHY-2020275, and the Heising-Simons Foundation (2017-228).
Y.W. is grateful for mentorship from his advisor Roberto Car, and acknowledges support from the DOE
Award DE-SC0017865. 

\bibliographystyle{apsrev4-1}
\bibliography{main.bib}
%========================================================================================

\appendix

\section{PROJECTION OPERATOR PRODUCT REDUCTION}
\label{sec:ProjectionReduction}
Current hardware topology requires the qubit or spin Hamiltonian to have two-local interactions.
For Eq. \ref{eqn:graylap}, fortunately, there is a known method \cite{PhysRevA.78.012320} to replace a product of $P_i^0$s to the projection operator of a single qubit while keeping the low-lying spectrum of the Hamiltonian, at the expense of adding ancillary qubits\footnote{The $\hat{q}_i$ in \cite{PhysRevA.78.012320} is  $P^1_i$ here.}.
Because each iteration of the recursion formula shares all but one projection operator with its predecessor, the ancillary bit cost is linear in $A$.
We include a brief discussion of the construction of the reduction.

To reduce the correction terms to 2-local, it is sufficient to reduce the product of projection operators to the projection operator of a single ancillary qubit.
If we can reduce two qubits to one, then a tree or chain of such reductions will suffice.
We construct the 2 to 1 reduction by adding  qubit $a$  along with a penalty contribution to the Hamiltonian.
To replace  $P_{i}^0 P_{j}^0$, for example, a penalty is added when  $P_a^0 \ne P_{i}^0 P_{j}^0$.
The penalty is shown for all  states of qubits $a$, $i$, and $j$ in  \tabref{tab:reduceprojection}.
\begin{table}[h]
\caption{\label{tab:reduceprojection}
A diagonal penalty Hamiltonian for reducing the product of two qubit projection operators $p_i^0$ and $p_j^0$ to a single qubit labeled $q$ that is 0 when the original qubits are both 0.   The $Q$ indicates a sufficient penalty to remove violators from the low lying states, not a specific value.   }
\begin{tabular}{cc|ccc|c}
\hline\hline
\textrm{a} & $P_a^0$ & \textrm{i} & \textrm{j} & $P_{i}^0 P_{j}^0$ & $H_{pen}$ \\
\hline\hline
0 & 1 & 0 & 0 & 1 & 0 \\
0 & 1 & 0 & 1 & 0 & Q \\
0 & 1 & 1 & 0 & 0 & Q \\
0 & 1 & 1 & 1 & 0 & Q \\
1 & 0 & 0 & 0 & 1 & Q \\
1 & 0 & 0 & 1 & 0 & 0 \\
1 & 0 & 1 & 0 & 0 & 0 \\
1 & 0 & 1 & 1 & 0 & 0 \\
\end{tabular}
\end{table}
\begin{figure}[h]
\centering
\includegraphics[scale=0.36 ]{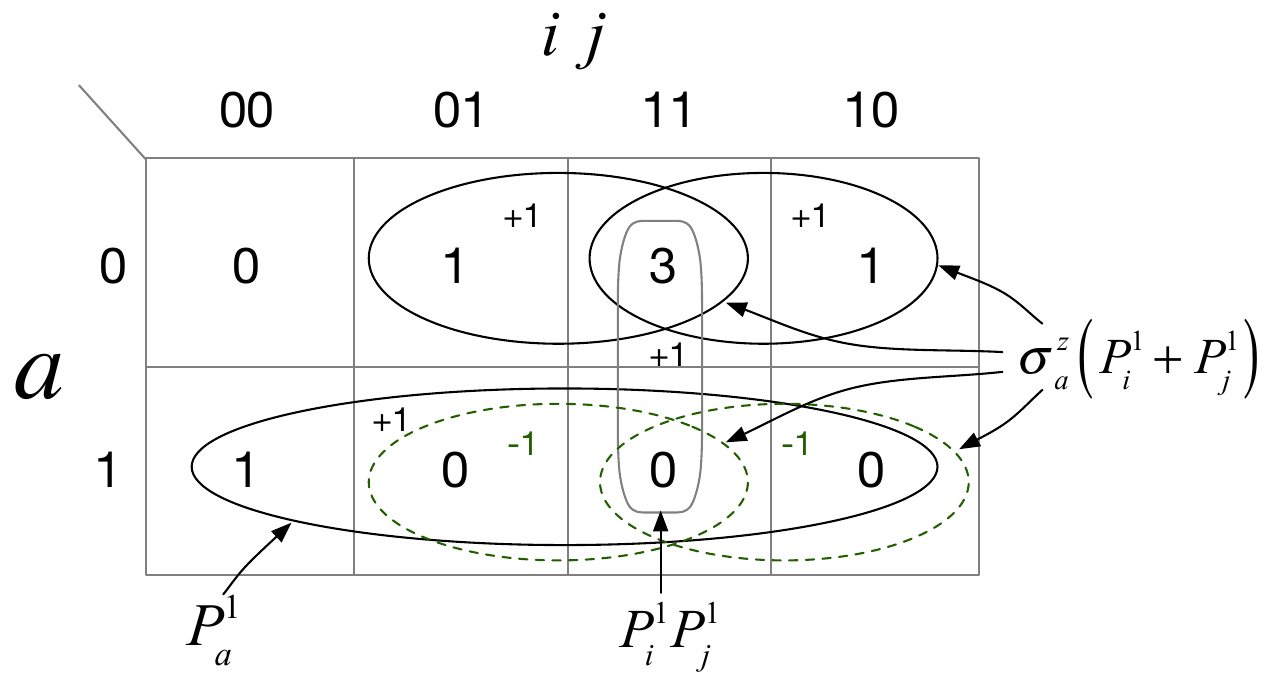}
\caption{ Penalties from \tabref{tab:reduceprojection} appear here as non-zero values in the center of the cells, which will
be multiplied by Q.
Contributions are shown as ovals with +1 or -1 indicating the coefficient for the contribution.    The sum of such
values for all ovals covering the center of a cell totals to the penalty value of the cell.
When restricted to the first row, $P_{i}^1 + P_{j}^1$ generates the two ovals in the row.
The same contribution with the opposite phase is needed in the bottom row, shown with green dashed ovals.
$\sigma_a^z \left(P_{i}^1 + P_{j}^1\right)$ generates all 4 contributions.    A value is 1 is added in the bottom
row with $P_a^1$, leaving only cell 1/11 needing a correction, which is provided by $P_{i}^1 P_{j}^1$.
This process leaves cell 0/11 with a larger than needed but acceptable penalty of 3Q. }
\label{fig:PenKarnaugh}
\end{figure}
We use a Karnaugh map, shown in \figref{fig:PenKarnaugh}, to visualize the adjacencies and assist in minimizing the implementation of the penalty Hamiltonian.

The resulting penalty Hamiltonian is
\begin{equation}
H_{pen} = Q  \left[P_a^1 + \sigma_a^z \left( P_{i}^1 + P_{j}^1\right) + P_{i}^1 P_{j}^1 \right].
\end{equation}
All pieces of this penalty contribution are 2-local.

To reduce a collection of projector products to 2-local, an efficient heuristic is to rank projector pairs by the number of existing products in which they appear.
Then, the highest-ranked such pair is processed, producing a new qubit, and the projector on the new qubit
replaces the pair in every product in which it appears, repeating the process until completion.
This process is a well-known heuristic for reducing collections of multi-input boolean and gates.
The pair's tree height can also be included as a negative contribution in the ranking to avoid long chains of ancillary qubits.

%========================================================================================
\section{REVIEW OF THE ORTHOGONAL FUNCTIONS}
\label{sec:math_review}
%========================================================================================

\subsection{Walsh series}

The tensor product space of Pauli $\sz$ matrices form a complete basis for real valued functions and is analogous to a digitized Fourier series expansion. This can be understood by recognizing that there is a one-to-one mapping of the $\sz$ tensor products to the Walsh series which we will discuss in this section.

\subsubsection{Walsh and Rademacher functions}
Before defining the Walsh functions, let us first introduce the Rademacher functions $R_n$, which are the basic building blocks of the basis states. The functions $R_n$ are defined as
\begin{align}
  R_n(t) = \sign \sin(2^{n+1} \pi t),
\end{align}
where $t$ spans the unit interval, and $n$ is the set of natural numbers starting from zero. We can immediately interpret $R_n$ as the set of periodic step functions with integer frequencies as enforced by periodic boundary conditions.

In an encoding $G$, the Walsh functions are constructed from the Rademacher functions such that
\begin{align}
    W^G_0 \equiv & 1\\
    W^G_n =& \Pi_{n_i} R_{n_i},
\end{align}
where the set $\{n_i\}$ is composed of the positional indices of the non-zero bits in $G(n)$.
The bit-string representation of $n$ is read from left to right.
For example, if $G(n) = 110$, then $\{n_i\} = \{0,1\}$, and the corresponding Walsh function is $R_0(t)R_1(t)$.
By definition, $W^G_0$ corresponds to the zero-frequency mode in all encodings.
For a more concrete discussion, we list the first $2^3$ integers in binary, BRGC, sequency, and Hamming distance 2 Gray (H2GC) encoding

\begin{align}
    \label{eq:bit_order_label}
&\begin{pmatrix}0\\1\\2\\3\\4\\5\\6\\7 \end{pmatrix} & \xrightarrow[\textrm{rep}]{\text{bin}}
    \begin{pmatrix} 000 \\ 001 \\ 010 \\ 011 \\ 100 \\ 101 \\ 110 \\ 111 \end{pmatrix} \
    \begin{pmatrix} 000 \\ 001 \\ 011 \\ 010 \\ 110 \\ 111 \\ 101 \\ 100 \end{pmatrix} \
    \begin{pmatrix} 000 \\ 100 \\ 110 \\ 010 \\ 011 \\ 111 \\ 101 \\ 001 \end{pmatrix} \
    \begin{pmatrix} 0000 \\ 0001 \\ 0011 \\ 0111 \\ 1111 \\ 1110 \\ 1100 \\ 0100 \end{pmatrix} \\
 &\ \ \textrm{int} & \textrm{bin} \quad\quad \ \textrm{brgc} \quad\quad \textrm{seq}_3\quad \ \ \ \textrm{h2gc}\ \nonumber
\end{align}

The binary order is also called the Hadamard order in the literature, the BRGC order follows from Gray code discussed in Sec.~\ref{sec:graycode}, sequency order is also called the Walsh order in the literature and is just the reflection of the BRGC order for a given number of bits, and finally the H2GC sequence is discussed in Sec.~\ref{sec:Laplacian:Dist2Gray} and is used to encode the Laplacian with the transverse Ising model Hamiltonian.

\subsubsection{Binary order}
The Walsh functions $W^{\textrm{bin}}_n$ in binary order are denoted by a superscript $\textrm{bin}$. Following Eq.~(\ref{eq:bit_order_label}), we give the first three Walsh functions in binary order to illustrate the construction
\begin{align}
    1\rightarrow 001 && \therefore W_1^{\textrm{bin}} = & R_2,\nonumber \\
    2\rightarrow 010 && \therefore W_2^{\textrm{bin}} = & R_1,\nonumber \\
    3\rightarrow 011 && \therefore W_3^{\textrm{bin}} = & R_1 R_2. \nonumber
\end{align}

\subsubsection{Binary reflected Gray order}
An alternative way to order the Walsh functions is to map the sequence to BRGC, and is the computational ordering for the $XZ$-model mapping of the Schr\"odinger equation.
We use the notation $W^{\textrm{brgc}}_n$ to denote the Gray ordered Walsh function. Following Eq.~(\ref{eq:bit_order_label}), the first three Walsh functions in Gray order are
\begin{align}
    1\rightarrow 001 && \therefore W_1^{\textrm{brgc}} = &R_2, \nonumber \\
    2\rightarrow 011 && \therefore W_2^{\textrm{brgc}} = &R_1 R_2, \nonumber \\
    3\rightarrow 010 && \therefore W_3^{\textrm{brgc}} = &R_1. \nonumber
\end{align}

\subsubsection{Sequency order}
The sequency order is analogous to the Fourier series mode expansion, and was the version originally employed by Walsh~\cite{Walsh1923}. In this order, each function has one more zero crossing than the previous function and the set alternates between even and odd functions sequentially. From this perspective, it is very similar to the Fourier series and the concept of frequency is replaced by senquency. The list of sequency bit-strings are obtained by performing a bit-reversal on the BRGC bit-strings. Due to bit-reversal, the sequency order mapping is dependent on the total size of the system $A$.
We use the notation $W^{\textrm{seq}_A}_n$ to denote the sequency ordered Walsh functions for an $A$ (qu)bit system. Following Eq.~(\ref{eq:bit_order_label}), the first three Walsh functions in sequency order for a 3-bit system are
\begin{align}
    1\rightarrow 100 && \therefore W_1^{\textrm{seq}_3} = &R_0, \nonumber \\
    2\rightarrow 110 && \therefore W_2^{\textrm{seq}_3} = &R_0 R_1, \nonumber \\
    3\rightarrow 010 && \therefore W_3^{\textrm{seq}_3} = &R_1. \nonumber
\end{align}

As a result, low-mode expansions can be computed successively one contribution at a time given the above sequency order. In Sec.~\ref{sec:potential_decomposition} we suggest using a combination of coarse graining and the Fast Walsh Transform (similar to the Fourier version) in order to gain a substantial computational speed up when series expanding arbitrary real functions. Therefore, this discussion of the sequency ordering is meant to give better intuition for the Walsh series, and are important when discussing the series expansion for potentials.
The Walsh functions in the sequency order are also given by the rows of the Hadamard matrix,
\begin{align}
      \label{eq:Hmatrix}
  H(2^k)=&\begin{pmatrix} H(2^{k-1})  && H(2^{k-1}) \\ H(2^{k-1}) && -H(2^{k-1})\end{pmatrix}, \\
  H(2^1)=&\begin{pmatrix} 1  && 1 \\ 1 && -1\end{pmatrix}. \nonumber
\end{align}
Then, $W_n^{\textrm{bin}} = n^{\text{th}}\ \text{row of}\ H(2^k)$.

\subsubsection{Mapping to the Pauli basis}
\begin{figure}[ht]
\centering
\includegraphics[scale=0.5 ]{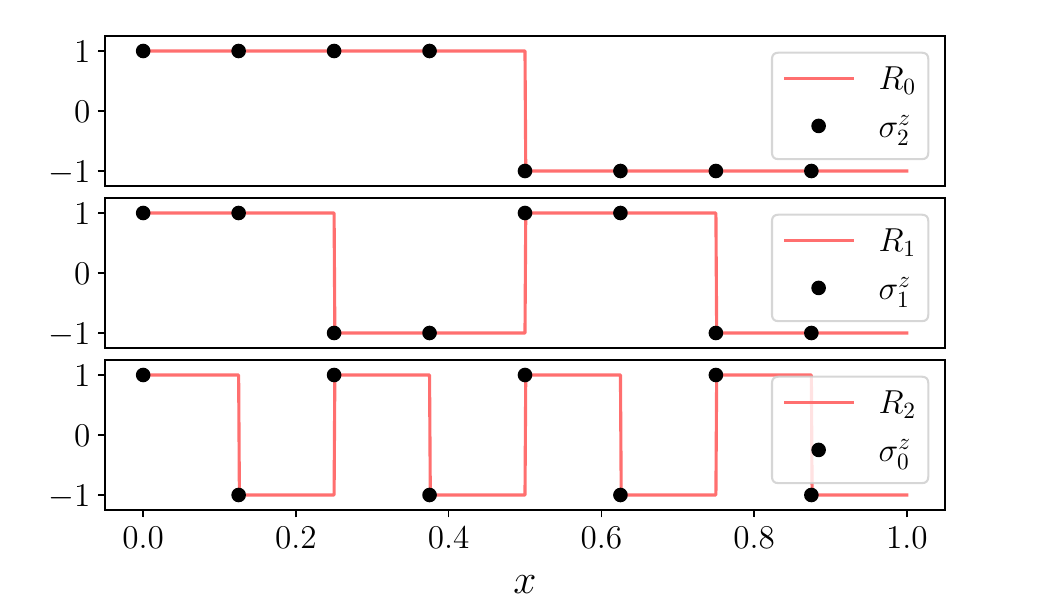}
\caption{The first 3 Rademacher functions.}
\label{fig:Rademacher}
\end{figure}
For a system of $A$ qubits, the first $2^A$ Rademacher functions have an exact mapping to the diagonal of the $1-$local $\sigma^z$ Hamiltonian, $R_n \rightarrow \sigma^z_{A-1-n}$. The first three Rademacher functions are shown in Fig.~\ref{fig:Rademacher} along with the corresponding $1-$local Hamiltonian for a system of 3 qubits.

It follows immediately that given a system of $A$ qubits, the set of $2^A$, $k$-local Ising-like Hamiltonians are bijectively mapped to the first $2^A$ Walsh functions.
For example, in a system of 3 qubits, the $n=4$ Walsh function in binary order is given by
\begin{align}
  W^{\textrm{bin}}_4 = R_2 \rightarrow \sigma^z_{0} = \mathbb{1}\otimes \mathbb{1} \otimes \sz
\end{align}
and in Gray order as
\begin{align}
  W^{\textrm{brgc}}_4 = R_0 R_1 \rightarrow \sigma^z_{2} \sigma^z_{1} =  \sz \otimes \sz \otimes \id
\end{align}
and in Walsh order as
\begin{align}
    W^{\textrm{seq}_3}_4 = R_1 R_2 \rightarrow \sigma^z_1 \sigma^z_0 = \id \otimes \sz \otimes \sz.
\end{align}

In general, given a binary representation for an integer $n$, the 1s and 0s map respectively to tensor products of $\sigma^z$ and $\mathbb{1}$. Fig.~\ref{fig:Walsh_sequency_basis} shows the first 7 Walsh functions in sequency order and highlights the connection to sine and cosine functions with increasing frequency.

\begin{figure}[ht]
\centering
\includegraphics[scale=0.5 ]{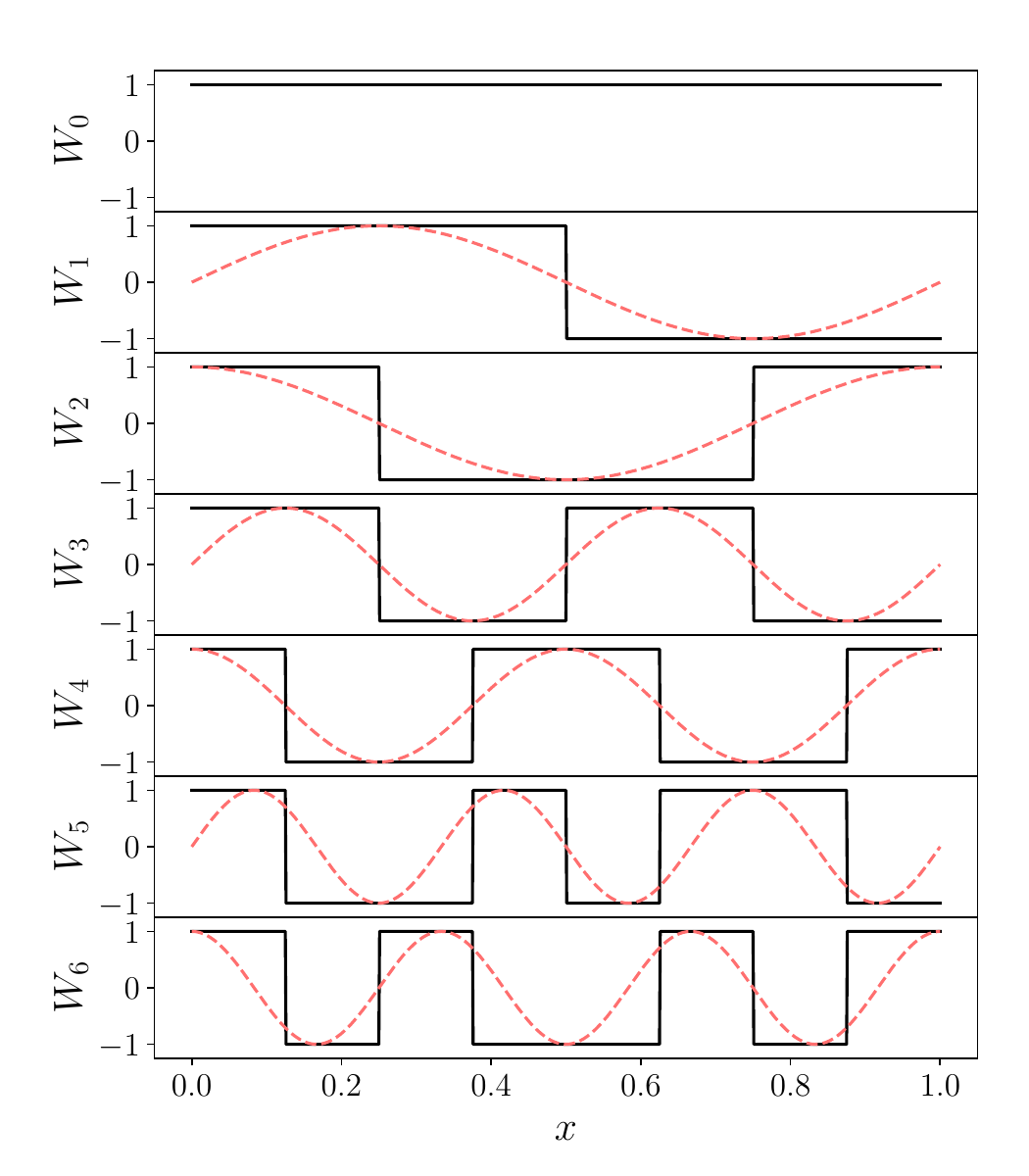}
\caption{The first 7 Walsh functions in sequency order.}
\label{fig:Walsh_sequency_basis}
\end{figure}

\subsection{Fast Walsh-Hadamard Transform}

In analogy with the Fourier series, the Walsh functions in a given order form an orthonormal basis for the vector space of functions defined on $[0,1]$.
As already touched upon, differently from the Fourier series, there are many versions of the Walsh series depending on how the functions are ordered and sequency most closely resembles the concept of frequency with each subsequent element in the series increasing the number of zero crossings by one.
The expansion can be performed through the inner product,
\begin{equation}
 \begin{split}
  f(x)=&\sum_n a_n W_n(x),\\
  a_n=& \int_0^1 f(x)W_n(x).\\
 \end{split}
\end{equation}
On a discretized domain of $N=2^A$ equally spaced grid points, the Walsh-Hadamard transform can be easily realised from Eq.~(\ref{eq:Hmatrix}),
\begin{equation}
 \begin{split}
  \vec{f}^{(W)}=\frac{1}{2^A}H\left(2^A\right)\vec{f}
 \end{split}
\end{equation}
where the real function, $\vec{f}=\{f(x_i)\}_{i=0}^{2^A-1}$, has been evaluated at the grid points.
This transformation requires $N^2$ operations, just like the Discrete Fourier Transform (DFT), and indeed is equivalent
to a multidimensional DFT of size $2^A$~\cite{Kunz1979}.
In practice, one opts for an efficient implementation like the Fast Fourier Transform (FFT)~\cite{CooleyTukey1965}. This is achieved by the Fast Walsh-Hadamard Transform (FWHT) which requires $N \log (N)$ operations. Through the decades, various fast algorithms have been developed, which automatically return the expansion in a given order. As an illustration, here we provide the decompostion in binary order for a sequence of $4$ grid points by matrix partioning techniques~\cite{Ahmed1970}, where
\begin{align}
  4 \begin{pmatrix} f^{(W,b)}(x_0) \\ f^{(W,b)}(x_1) \\ f^{(W,b)}(x_2) \\ f^{(W,b)}(x_3)\end{pmatrix} =&  H(4) \begin{pmatrix} f(x_0) \\ f(x_1) \\ f(x_2) \\ f(x_3)\end{pmatrix}\nonumber \\
  =& \begin{pmatrix} H(2)  && H(2) \\ H(2) && -H(2)\end{pmatrix} \begin{pmatrix} f(x_0) \\ f(x_1) \\ f(x_2) \\ f(x_3)\end{pmatrix} \nonumber
  \end{align}
partitions to
\begin{align}
  4 \begin{pmatrix} f^{(W,b)}(x_0) \\ f^{(W,b)}(x_1) \end{pmatrix}=&  H(2)  \begin{pmatrix} f_1(x_0) \\ f_1(x_1) \end{pmatrix} \nonumber \\
  =&  H(2)  \begin{pmatrix} f(x_0) + f(x_2) \\ f(x_1) + f(x_3)\end{pmatrix}, \nonumber \\
  4 \begin{pmatrix} f^{(W,b)}(x_2) \\ f^{(W,b)}(x_3) \end{pmatrix}=&  H(2)  \begin{pmatrix} f_1(x_2) \\ f_1(x_3) \end{pmatrix} \nonumber \\
  =&  H(2)  \begin{pmatrix} f(x_0) - f(x_2) \\ f(x_1) - f(x_3)\end{pmatrix} \nonumber
  \end{align}
which can be further partitioned into
\begin{align}
  4 f^{(W,b)}(x_0)= & f_2(x_0)= \left( f_1(x_0) + f_2(x_1) \right), \nonumber \\
  4 f^{(W,b)}(x_1)= & f_2(x_1)= \left( f_1(x_0) - f_2(x_1)  \right), \nonumber \\
  4 f^{(W,b)}(x_2)= & f_2(x_2)= \left( f_1(x_2) + f_2(x_3) \right), \nonumber \\
  4 f^{(W,b)}(x_3)= & f_2(x_3)= \left(  f_1(x_2) - f_2(x_3) \right). \nonumber
\end{align}

\section{DERIVATION OF EQ. \ref{eqn:graylap} IN TENSOR PRODUCT NOTATION}
\label{sec:alt_deriv}
We start from the recursion formula of $L^{(A,\text{bin})}$ in Eq. \ref{eq:L_recursion}:
\begin{equation}
  L^{(A,\text{bin})} = \id \otimes (L^{(A-1,\text{bin})}-C_{A-1}) + \sigma^x\otimes C_{A-1}
  \label{appeq:L}
\end{equation}
We follow the notation in Sec. \ref{sec:graycode}.
For an integer $N = 2^A$, the encoding function, $G_A$, of BRGC is a permutation of $(0,1,2, \cdots, N-1)$.
It is defined inductively.
$G_1 = (0, 1)$.
For $A>1$, the first half of $G_A$ is $G_{A-1}$, and the second half of $G_{A}$ is $G_{A-1}$ reversed in order and then added by $2^{A-1}$.
For example, $G_2$ is $(0, 1)$ concatenated with $(1+2, 0+2)$, which is $(0, 1, 3, 2)$.
In particular,
\begin{equation}
  \begin{split}
  &G_A(0) = 0 \\
  &G_A(2^A-1) = 2^{A-1} + G_A(0) = 2^{A-1}\\
  &G_A(2^{A-1}-1) = 2^{A-2} \\
  &G_A(2^{A-1}) = 2^{A-1} + G_A(2^{A-1}-1) = 2^{A-1} + 2^{A-2}
\end{split}
\label{eq:Gn}.
\end{equation}
Let the matrix transformation in Eq. \ref{eq:gray_transform} be denoted by $\G_A$:
\begin{equation}
  L^{(A, \text{brgc})} \equiv \G_A(L^{(A, \text{bin})}).
\end{equation}
To derive a formula for $L^{(A,\text{brgc})}$, we first note that $\G_A(\id\otimes M) = \id\otimes \G_{A-1}(M)$ if $M$ is invariant under the reflection permutation, $(A-2, A-3, \cdots, 0)$, and that $L^{(A-1,\text{bin})}$ and $C^{A-1}$ both enjoy this invariance.
Thus,
\begin{equation}
  L^{(A,\text{brgc})} = \id \otimes L^{(A-1,\text{brgc})} -\id  \otimes \G_{A-1}(C_{A-1}) + \G_A(\sx\otimes C_{A-1}).
\end{equation}
To compute $\G_{A}(C_{A})$, note that $(C_A)_{ij}$ is nonzero if $(i,j) = (0, 2^A-1)$ or $(2^A-1,0)$.
Thus, $\G_A(C_A)$ is nonzero at $(0, 2^{A-1})$ and $(2^{A-1}, 0)$.
For example, for $A=2$,
\begin{equation}
  \G_2(C_2) = \begin{bmatrix}
    0&0&1&0\\
    0&0&0&0\\
    1&0&0&0\\
    0&0&0&0\\
  \end{bmatrix} = \sx \otimes P^0
\end{equation}
where $P^0 = (1-\sz)/2$ is a $z$-projection matrix.
It is easy to see that, for general $A$,
\begin{equation}
  \G_A(C_A) = \sx \otimes (P^0)^{\otimes(A-1)}
\end{equation}
To compute $\G_A(\sx\otimes C_{A-1})$, we note that $(\sx\otimes C_{A-1})_{ij}$ is nonzero at $(i,j)$ = $(2^A-1,0)$, $(0,2^A-1)$, $(2^{A-1},2^{A-1}-1)$, and $(2^{A-1}-1,2^{A-1})$.
According to Eq. \ref{eq:Gn}, this means that $\G_A(\sx\otimes C_{A-1})$ is nonzero at $(2^{A-1},0)$, $(0, 2^{A-1})$, $(2^{A-1}+2^{A-2}, 2^{A-2})$, and $(2^{A-2}, 2^{A-1}+2^{A-2})$.
For example, for $A=3$,
\begin{equation}
  \G_3(\sx\otimes C_2) =
  \begin{bmatrix}
    0& 0& 0& 0& 1& 0& 0& 0 \\
    0& 0& 0& 0& 0& 0& 0& 0 \\
    0& 0& 0& 0& 0& 0& 1& 0 \\
    0& 0& 0& 0& 0& 0& 0& 0 \\
    1& 0& 0& 0& 0& 0& 0& 0 \\
    0& 0& 0& 0& 0& 0& 0& 0 \\
    0& 0& 1& 0& 0& 0& 0& 0 \\
    0& 0& 0& 0& 0& 0& 0& 0 \\
  \end{bmatrix} = \sx \otimes \id \otimes P^0.
\end{equation}
For general $n$, we see that
\begin{equation}
  \G_A(\sx\otimes C_{A-1}) = \sx \otimes \id \otimes (P^0)^{\otimes (A-2)}
\end{equation}
Thus, we obtain
\begin{equation}
  L^{(A,\text{brgc})} = \id\otimes L^{(A-1,\text{brgc})} + (\sx\otimes \id - \id \otimes \sx) \otimes (P^0)^{\otimes(A-2)},
\end{equation}
where the base case is $L_2 = \id\otimes\sx + \sx\otimes \id$.

\section{INTRODUCTION TO KARNAUGH MAPS}
\label{sec:KMap}
Karnaugh maps~\cite{karnaugh1953map} are a tool for visualizing binary hypercubes with dimension $\ge 3$.
A common use of them in classical boolean circuit design is as an aid in minimization of boolean functions as
a sum of products or a product of sums.
The cells in a Karnaugh map represent the corners of a binary hypercube in a way that makes it
easy to visually identify  sub-cubes of the complete hypercube.
\figref{fig:KMap} shows a Karnaugh map for a function of 4 variables.
Sub-cubes are important because they can also be specified as a boolean product
of boolean literals (a literal is a boolean variable or it's complement) for variables that that do not change in the sub-cube.
In the quantum computing context the parallel specification is  a product of projection operators on the qubits
who's values are constant in the sub-cube.
\begin{figure}[ht]
\centering
\includegraphics[scale=0.33 ]{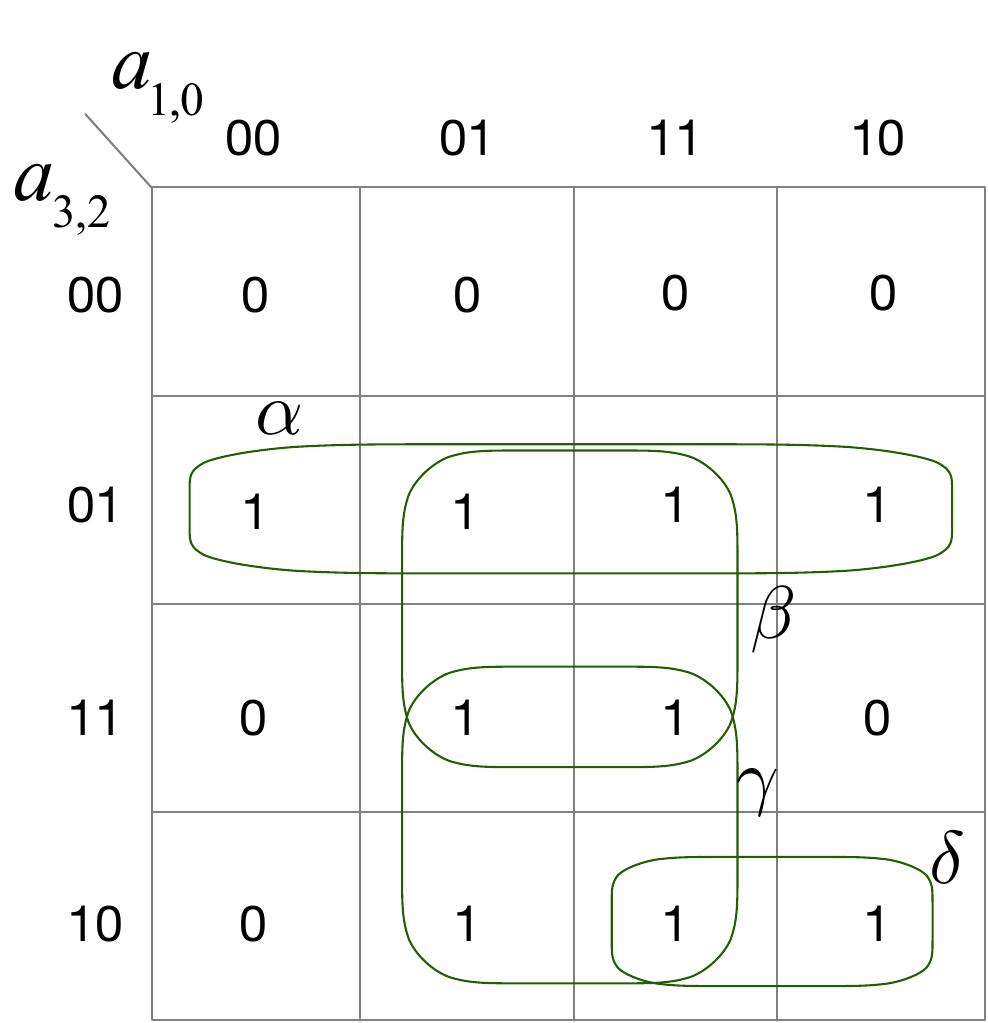}
\caption{A Karnaugh map on 4 input variables.   The 16 cells correspond to the corners of a dimension 4 binary hypercube, each of
which is associated with specific values of the input variables.
The input values can be read from the left of the row and the opt of the column the cell is in.
Each cell contains the boolean function value for the input state of the cell.
The greek letter labels have been added for the discussion are not normally part of
a Karnaugh map.    The left most 1 under the label $\alpha$ is in a cell with input values 0100.
Note that the column and row input values follow the sequence 00, 01, 11, 10, which has the property that visually adjacent cells differ in the value of only one input variable.
We also consider the right hand edge to be adjacent to the left hand edge and the bottom edge to be adjacent to the top edge.   }
\label{fig:KMap}
\end{figure}
In \figref{fig:KMap} there are 4 groups indicated with labels $\alpha$, $\beta$, $\gamma$, and $\delta$.   Each of these groups is a sub-cube of the full 4 variable hypercube.    We can write a simple product expression for each group.   Note that the bar over variables is a standard way to indicate a boolean complement.
\begin{equation}
\alpha = \bar{a}_3 a_2 ,\;
\beta = a_2 a_0 ,\;
\gamma = a_3 a_0 ,\;
\delta = a_3 \bar{a}_2 a_1
\end{equation}
\begin{figure}[h!]
\centering
\includegraphics[scale=0.25 ]{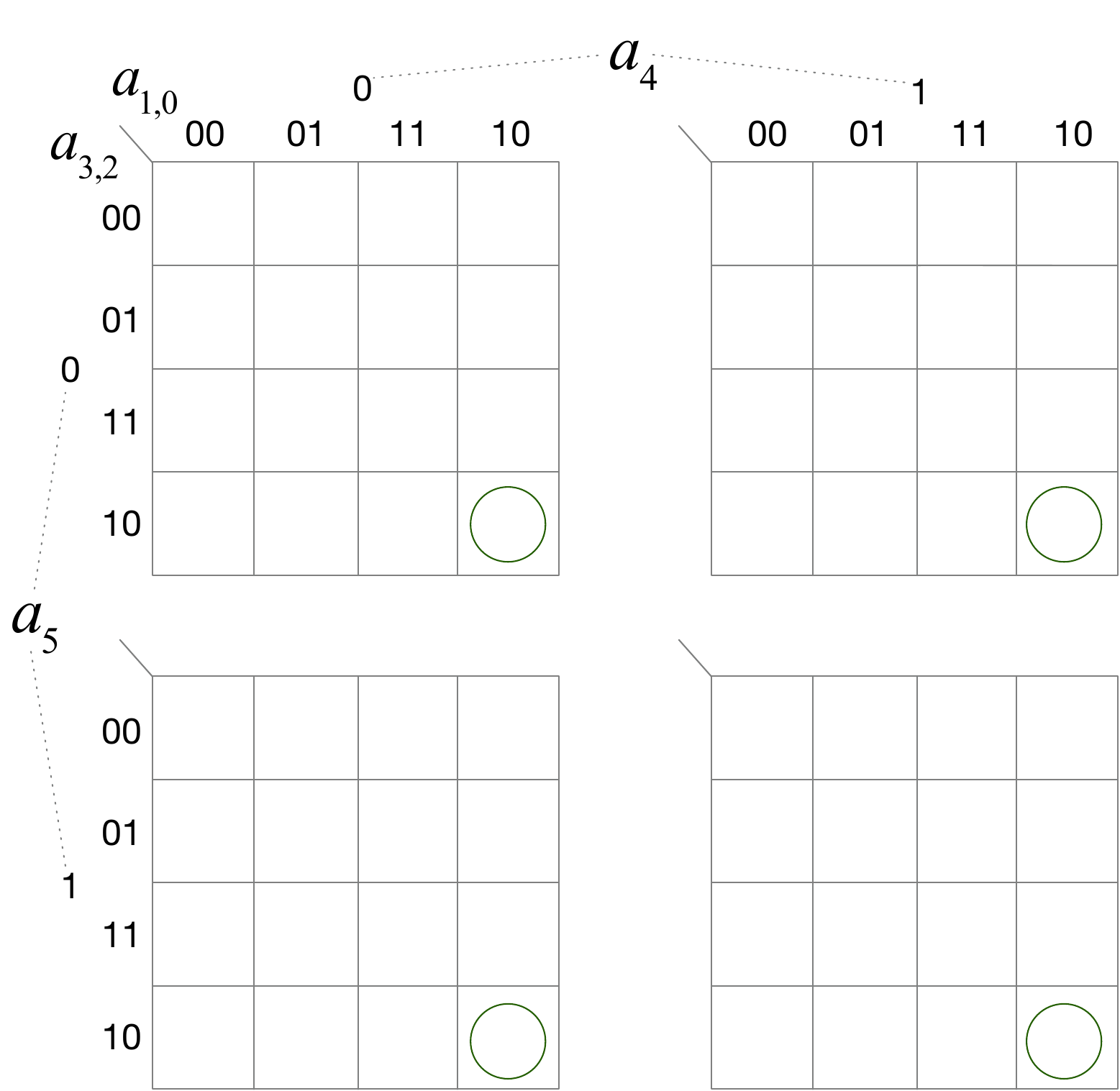}
\caption{A Karnaugh map on 6 input variables.   One should regard the sub-maps as separate pieces of paper stacked up in the order $a_{5,4}=00,01,11,10$, in clockwise order.   The top and bottom sheets are adjacent because of periodicity.    A cell is considered adjacent to the cell in the sheet above or below it in addition to adjacencies in the 4 variable map. As an example, a 2 variable sub-cube with 4 cells can be formed from the collection of the lower right cells of all four sub maps, with inputs $a_{3,2,1,0}=1010$ or $a_3 \bar{a}_2 a_1 \bar{a}_0$.}
\label{fig:KMap6}
\end{figure}

In using a Karnaugh map for optimization of a sum of products one first finds the set of maximal sub-cubes containing only 1s.  The indicated groups are maximial because removal of any variable from their product expression would enlarge them to include a cell with a 0.  Next, a subset
of the sub-cubes are selected such that every 1 is inside one of the selected sub-cubes.   In this example sub-cube $\beta$
is unnecessary because all the contained 1s are covered by sub-cubes $\alpha$ and $\gamma$, which is easily checked by examination of the Karnaugh map.    The complete boolean function $f$ is then
\begin{equation}
f = \bar{a}_3 a_2 + a_3 a_0 + a_3 \bar{a}_2 a_1
\end{equation}
In our quantum computing application to penalties the sum operation is a numeric sum instead of a boolean sum (also known as a boolean or).   For penalties however,
we do not care about the exact penalty value and two terms of the same sign may both contribute without harm, giving the same character as the boolean sum.

The simple Karnaugh map in \figref{fig:KMap} can be extended to more variables in a hierarchical way.
\figref{fig:KMap6} shows a 6 input map, which can be thought of as a 2 variable map with each cell containing a 4 variable map.
An 8 input map can be constructed as a 4 variable map with each cell being in turn a 4 variable map.

\end{document}